\documentclass[12pt]{article}

\newcommand{\newsection}{\setcounter{equation}{0}\section}

\newcommand{\eqn}[1]{(\ref{#1})}
\def\appendix#1{\addtocounter{section}{1}\setcounter{equation}{0}
\renewcommand{\thesection}{\Alph{section}}
\section*{Appendix~~\thesection\protect\indent \parbox[t]{11.715cm} {#1}}
\addcontentsline{toc}{section}{Appendix \thesection\ \ \ #1} }
\newcommand{\complex}{{\bb C}} 
\newcommand{\complexs}{{\bbs C}} 
\newcommand{\zed}{{\bb Z}} 
\newcommand{\real}{{\bb R}} 
\newcommand{\reals}{{\bbs R}} 
\newcommand{\zeds}{{\bbs Z}} 
\newcommand{\mat}{{\bb M}} 
\newcommand{\NO}{\,\mbox{$\circ\atop\circ$}\,} 
\newcommand{\id}{{1\!\!1}} 
\def\alg{{\cal A}}
\def\hil{{\cal H}}

\font\mybb=msbm10 at 12pt
\def\bb#1{\hbox{\mybb#1}}
\font\mybbs=msbm10 at 9pt
\def\bbs#1{\hbox{\mybbs#1}}

\def\nn{\nonumber}
\newcommand{\tr}[1]{\:{\rm tr}\,#1}
\newcommand{\Tr}[1]{\:{\rm Tr}\,#1}
\def\e{{\,\rm e}\,}

\newcommand{\non}{\nonumber \\*}

\def\be{\begin{equation}}
\def\ee{\end{equation}}
\def\bea{\begin{eqnarray}}
\def\eea{\end{eqnarray}}
\def\bd{\begin{displaymath}}
\def\ed{\end{displaymath}}

\def\dd{{\rm d}}

\makeatletter
\newdimen\normalarrayskip              
\newdimen\minarrayskip                 
\normalarrayskip\baselineskip
\minarrayskip\jot
\newif\ifold             \oldtrue            \def\new{\oldfalse}
\def\arraymode{\ifold\relax\else\displaystyle\fi} 
\def\@arrayskip{\ifold\baselineskip\z@\lineskip\z@
     \else
     \baselineskip\minarrayskip\lineskip2\minarrayskip\fi}
\def\@arrayclassz{\ifcase \@lastchclass \@acolampacol \or
\@ampacol \or \or \or \@addamp \or
   \@acolampacol \or \@firstampfalse \@acol \fi
\edef\@preamble{\@preamble
  \ifcase \@chnum
     \hfil$\relax\arraymode\@sharp$\hfil
     \or $\relax\arraymode\@sharp$\hfil
     \or \hfil$\relax\arraymode\@sharp$\fi}}
\def\@array[#1]#2{\setbox\@arstrutbox=\hbox{\vrule
     height\arraystretch \ht\strutbox
     depth\arraystretch \dp\strutbox
     width\z@}\@mkpream{#2}\edef\@preamble{\halign \noexpand\@halignto
\bgroup \tabskip\z@ \@arstrut \@preamble \tabskip\z@ \cr}%
\let\@startpbox\@@startpbox \let\@endpbox\@@endpbox
  \if #1t\vtop \else \if#1b\vbox \else \vcenter \fi\fi
  \bgroup \let\par\relax
  \let\@sharp##\let\protect\relax
  \@arrayskip\@preamble}
\makeatother

\newcommand{\beq}{\begin{eqnarray}}
\newcommand{\eeq}{\end{eqnarray}}

\newcommand{\hk}{{\rm K}}

\newcommand{\del}{\partial}

\begin{document}
\begin{titlepage}
\begin{flushright}

\baselineskip=12pt
DSF--19--01\\ HWM--01--25\\ EMPG--01--10\\ hep--th/0107115\\
\hfill{ }\\
July 2001
\end{flushright}

\vspace{0.5 cm}

\begin{center}

\baselineskip=24pt

{\Large\bf Geometry of the Gauge Algebra in \\ Noncommutative Yang-Mills
Theory}

\baselineskip=14pt

\vspace{1cm}

{ {\bf F.\ Lizzi} $^{a}$, {\bf R.J.~Szabo} $^{b}$ and {\bf A.~Zampini} $^{a}$}
\\[6mm]
$^a$ {\it Dipartimento di Scienze Fisiche, Universit\`{a} di Napoli {\sl
Federico II}\\ and {\it INFN, Sezione di Napoli}\\ Monte S.~Angelo, Via
Cintia, 80126 Napoli, Italy}\\ {\tt fedele.lizzi , alessandro.zampini
@na.infn.it}
\\[6mm]
$^b$ {\it Department of Mathematics, Heriot-Watt University\\ Riccarton,
Edinburgh EH14 4AS, Scotland}\\  {\tt R.J.Szabo@ma.hw.ac.uk}
\\[10mm]

\end{center}

\vskip 3 cm

\begin{abstract}

\baselineskip=12pt

A detailed description of the infinite-dimensional Lie algebra of $\star$-gauge
transformations in noncommutative Yang-Mills theory is presented. Various
descriptions of this algebra are given in terms of inner automorphisms of the
underlying deformed algebra of functions on spacetime, of deformed symplectic
diffeomorphisms, of the infinite unitary Lie algebra $u(\infty)$, and of the
$C^*$-algebra of compact operators on a quantum mechanical Hilbert space. The
spacetime and string interpretations are also elucidated.

\end{abstract}

\end{titlepage}

\newpage

\setcounter{page}{1}

\newsection{Introduction}

One of the most interesting aspects of Yang-Mills theory on a noncommutative
space~\cite{book}--\cite{munich} is its extended gauge symmetry. This symmetry
group mixes the internal gauge degrees of freedom with the geometrical degrees
of freedom in spacetime. There are several interesting consequences of this
feature. The most striking one is that noncommutative gauge theory does not
contain any local observables in the usual sense, because ordinary traces
must be accompanied by an integration over the space in order to render
operators gauge invariant. On the other hand, the noncommutative gauge
symmetry seemingly allows for a larger class of observables. In addition to
the usual loop observables, there are gauge invariant Wilson line operators
associated with open contours which may be thought of as carrying a
non-vanishing momentum~\cite{IIKK}--\cite{AMNS2}. From these objects it is
possible to construct gauge invariant operators which carry definite momentum,
and which reduce to the usual local gauge invariant operators of ordinary gauge
field theory
in the commutative limit~\cite{DHI}. In the D-brane picture, these gauge
invariant operators naturally couple to general background supergravity
fields~\cite{SUGRA}.

The construction of gauge invariant observables associated with open Wilson
lines relies on the translational invariance that persists in noncommutative
gauge theories~\cite{AMNS2,DHI}. The key observation is that translations in
the noncommutative directions are equivalent (up to global symmetry
transformations) to gauge transformations, i.e. they can be realized via
conjugation by unitary elements of the gauge symmetry group. The only other
theory that possesses such a property is general relativity, although in that
case it is not so
straightforward to write down a set of non-local invariant observables. In this
paper we shall explore this feature further and determine to what extent
noncommutative Yang-Mills theory may be regarded as a model of general
relativity, i.e. as a theory whose local gauge symmetry includes general
coordinate transformations.

There are several hints that noncommutative gauge theories on {\it flat}
space naturally possess general covariance.
\begin{itemize}
\item{In string theory, the unitary group of a closed string vertex operator
algebra contains generic reparametrizations of the target space
coordinates~\cite{LLS}. From this fact it is possible to realize general
covariance as a gauge symmetry by using the chiral structure of the closed
strings.}
\item{Noncommutative gauge theories arise most naturally in string theory. In
particular, they are typically defined on closed string backgrounds of the form
${\cal M}\times\real^d$, where $\cal M$ is the commuting worldvolume of a
D-brane. The effect of turning on a non-degenerate Neveu-Schwarz $B$-field
along the transverse space $\real^d$ can be absorbed, in a particular
low-energy limit, into a description of the dynamics in terms of noncommutative
gauge theory~\cite{CDS,SW}. As such, these field theories possess many stringy
structures, but within a much simpler setting. For example, the one-loop
long-ranged potential particular to noncommutative Yang-Mills theory can be
identified with the gravitational interaction in Type IIB superstring
theory~\cite{AIIKKT,IKKKK}.}
\item{Certain large $N$ matrix models provide a concrete, non-perturbative
definition of noncommutative Yang-Mills theory~\cite{AMNS1,AIIKKT}, and these
discrete models have a $U(\infty)$ symmetry group which may be identified
with the group of area-preserving diffeomorphisms in
two-dimensions~\cite{FairlieZachos}. It is natural then to expect that
noncommutative gauge symmetries contain at least the subgroup of symplectic
diffeomorphisms of the spacetime.}
\item{In the large $N$ dual supergravity description of noncommutative
Yang-Mills theory in four dimensions with 16 supersymmetries, there is a
massless bound state which gives rise to the Newtonian gravitational force
law~\cite{IIKK2}.}
\item{Via certain dimensional reduction techniques, the translational symmetry
of noncommutative Yang-Mills theory can be gauged to induce a field theory
which contains as special limits some gauge models of
gravitation~\cite{LSred}.}
\end{itemize}
Other indications that noncommutativity implies general covariance can be found
in~\cite{Sochichiu,AIKO}. It therefore seems likely that noncommutative
Yang-Mills theory contains both gauge theory and gravitation, and so is a good
candidate for a unified and potentially renormalizable theory of the
fundamental interactions including gravity.

To investigate these features further, in this paper we will give a detailed,
analytic description of the Lie algebra of local noncommutative gauge
transformations. This is particularly important for the interpretation of
noncommutative Yang-Mills theory as some sort of gauge model of spacetime
symmetries, because it admits a dual interpretation as ordinary Yang-Mills
theory with this extended, infinite-dimensional gauge symmetry
algebra~\cite{Sheikh}. At the same time this analysis will generally sharpen
the present understanding of the local structure of the noncommutative symmetry
groups underlying these gauge theories. We shall see in fact that many of the
non-local, stringy features of noncommutative field theories are captured by
the geometrical properties of their symmetry groups. We will find that this
infinite dimensional algebra is a deformation of the Lie algebra of symplectic
transformations of the space in a very precise way, the symplectic structure
being given by the noncommutativity parameters. A similar sort of geometric
description of the noncommutative gauge algebra has been given recently
in~\cite{NCFTrev}. We will also discuss the connections between this algebra
and the infinite unitary Lie algebra $u(\infty)$ which is the natural symmetry
algebra from the point of view of the twisted reduced models describing
noncommutative gauge theory. As we shall see, this description agrees with the
recent proposal in~\cite{Harvey}, where global aspects of the noncommutative
gauge group are described. Some other properties of the noncommutative gauge
algebra have been recently discussed in~\cite{CPST1}.

We will start in the next section by describing the relations between gauge
transformations and the inner automorphisms of the deformed algebra of
functions on spacetime. For this, we shall define the noncommutative gauge
symmetry by its finite representation on fundamental matter fields (similar
arguments appear in~\cite{GN}) and rigorously derive from this the
corresponding infinitesimal transformation rules. A corollary of this
derivation is that the fundamental representation is only irreducible in the
Schr\"odinger polarization of the noncommutative function algebra. This fact
immediately implies the description of~\cite{Harvey} in terms of the algebra of
compact operators acting on a single-particle, quantum mechanical Hilbert
space, and it also agrees with the recent observations of~\cite{CPST1}
concerning the irreducibility of the matrix representations of the
noncommutative gauge algebra.

In section~3 we explicitly describe properties of the gauge Lie algebra, and
calculate its structure constants. This somewhat technical analysis is carried
out in two different bases, and the relations between the two representations
are described. We point out various physical characteristics of the algebras
that we derive. In section~4 we will first describe how the gauge Lie algebra
is a deformation of the Poisson-Lie algebra of symplectic diffeomorphisms, to
which it reduces at the leading non-trivial order in the commutative limit. We
then describe the intimate relationships between this algebra and the
infinite-dimensional Lie algebra $u(\infty)$. This leads explicitly to a
description of the gauge algebra in terms of compact operators, as
in~\cite{Harvey}, and lends some insight into the relationships between these
gauge models, reduced models, and membrane physics. The relationships between
such geometric algebras and $u(\infty)$ have also been analysed recently
in~\cite{Merk}.

In section~5 we then describe various subalgebras of the gauge algebra, both
finite and infinite dimensional, and interpret them as geometric symmetry
transformations of the spacetime by examining the corresponding gauge
transformations of fields that they induce. In section~6 we briefly present
some applications of the technical formalism given in most of the paper. We
discuss some more algebraic aspects of the representation of global spacetime
coordinate transformations as gauge symmetries, and exemplify the fact that not
all diffeomorphisms are realizable as inner automorphisms, but rather only the
symplectic ones. We also comment on how the description of the gauge algebra in
terms of compact operators is particularly well-suited to describe some aspects
of solitons in noncommutative field theory, although a more thorough
investigation requires dealing with topological aspects of the noncommutative
gauge symmetry group (as in~\cite{Harvey}) which lies outside the scope of this
paper whose aim is to concentrate on the geometrical properties of the {\it
local} gauge group. Finally, in section~7 we close with some concluding
remarks.

\newsection{Inner Automorphisms in Noncommutative Gauge Theory}

In this section we will introduce some relevant concepts that will be used
throughout this paper. We start by introducing Yang-Mills theory based on a
noncommutative function algebra, and then describe the relationship between
the automorphisms of this algebra and noncommutative gauge transformations.

\subsection{Gauge Theory on Noncommutative Space}

In the simplest setting, a noncommutative space is defined as a space whose
coordinates $x_i$, $i=1,\dots,d$, generate the Lie algebra
\be
[x_i,x_j]=i\,\theta_{ij} \ ,
\label{commx}\ee
where $\theta_{ij}$ is a constant antisymmetric real-valued matrix of
length dimension 2. The algebra ${\cal A}_\theta$ of functions on this
space is the algebra generated by the $x_i$,
\beq
\alg_\theta={\cal S}\left(\real^d\right)\,/\,{\cal R}_\theta \ ,
\label{alggenxi}\eeq
where ${\cal R}_\theta$ stands for the commutation relations (\ref{commx}), and
${\cal S}(\real^d)$ denotes an appropriate Schwartz space of functions on
$\real^d\to\complex$ of rapid decrease at infinity which we regard as a
subspace in the closure of the ring $\complex[[x_1,\dots,x_d]]$ of formal power
series in the $x_i$. Later on we will be interested in appropriate
$C^*$-completions of (\ref{alggenxi}). The algebra $\alg_\theta$ can be
described as a deformation of the algebra $\alg_0$ of ordinary, continuous
functions on $\real^d\to\complex$. In the following we will also restrict the
functions of $\alg_0$ to those which lie in the appropriate (dense) Schwartz
space ${\cal S}(\real^d)$. The Banach norm on $\alg_0$ is the usual
$L^\infty$-norm,
\beq
\|f\|_\infty^{~}=\sup_{x\in\reals^d}\,\Bigl|f(x)\Bigr| \ , ~~ f\in\alg_0 \ .
\label{supnorm}\eeq
For simplicity, we will assume in this paper that the (Euclidean) spacetime
dimension $d$ is even and that $\theta_{ij}$ is of maximal rank. Otherwise,
the algebra $\alg_\theta$ has a non-trivial centre which we can quotient
out to effectively induce a non-degenerate deformation matrix.
Geometrically, this operation corresponds to the identification of an
ordinary, commutative subspace of noncommutative $\real^d$.

The deformation is described by the Groenewold-Moyal
$\star$-product~\cite{Groenewold,Moyal}
\be
(f\star g)(x)=\left.\e^{\frac i2\,\theta_{ij}\,
\partial_{\xi_i}\partial_{\eta_j}}\,
f(\xi)g(\eta)\right|_{\xi=\eta=x} \ , ~~ f,g\in\alg_0 \ ,
\label{defstar}\ee
so that when $\theta_{ij}=0$ the product reduces to the ordinary pointwise
multiplication of functions, while at higher orders in $\theta_{ij}$
derivatives of the functions appear. In this sense the $\star $-product is
non-local. By integrating (\ref{defstar}) by parts over $\real^d$ we see that
it possesses the cyclic integration property
\beq
\int\dd^dx~\Bigl(f_1\star\cdots\star f_n\Bigr)(x)=
\int\dd^dx~\Bigl(f_{\pi(1)}\star\cdots\star f_{\pi(n)}\Bigr)(x)
\label{starint}\eeq
for any collection of functions $f_1,\dots,f_n\in\alg_0$ and any cyclic
permutation $\pi\in S_n$.

Just as in the commutative case, we can introduce a linear
derivation $\partial_i$ of the algebra $\alg_\theta$ by defining
\beq
\del_i\,x_j=\delta_{ij} \ ,
\label{derivdef}\eeq
and then extending it to all of ${\cal A}_\theta$ using the usual Leibnitz
rule. The derivation $\partial_i$ can be represented as the $\star$-commutator
with an element of $\complex[[x_1,\dots,x_d]]$,
\be
\partial_if=-i\,\theta_i^{~j}\,[x_j,f]^{~}_\star \ ,
\label{partialinner}\ee
where
\beq
\theta_i^{~k}\,\theta_{kj}=\delta_{ij} \ ,
\label{thetainv}\eeq
and the Moyal bracket of two functions is given by
\be
[f,g]^{~}_\star(x)=(f\star g)(x)-(g\star
f)(x)=\left.2i\,\sin\left(\mbox{$\frac12$}
\,\theta_{ij}\,\partial_{\xi_i}\partial_{\eta_j}\right)\,
f(\xi)g(\eta)\right|_{\xi=\eta=x} \ .
\label{starcomm}\ee
The $\star$-commutator (\ref{starcomm}) satisfies the Jacobi identity and the
Leibnitz rule. Note that $\partial_i$ defines an inner derivation of the ring
$\complex[[x_1,\dots,x_d]]/{\cal R}_\theta$ but not of the algebra
$\alg_\theta$.

The action for $U(N)$ noncommutative Yang-Mills theory is given by
\be
S=\frac14\,\int\dd^dx~\tr\Bigl(F_{ij}(x)\star F^{ij}(x)\Bigr) \ ,
\label{NCYM}\ee
where
\be
F_{ij}=i\,[\nabla_i,\nabla_j]^{~}_\star=
\partial_iA_j-\partial_jA_i-i\,[A_i,A_j]^{~}_\star
\label{NCfieldstrength}\ee
is the noncommutative field strength tensor, and
\be
\nabla_i=\partial_i-i\,A_i
\label{gaugecovderiv}\ee
is the gauge connection. The gauge field $A_i(x)$ is a Hermitian element of the
algebra $\mat_N(\alg_\theta)=\alg_\theta\otimes\mat_N(\complex)$, where
$\mat_N(\complex)$ is the elementary $C^*$-algebra of $N\times N$ matrices and
the multiplication in $\mat_N(\alg_\theta)$ is the tensor product of the
$\star$-product (\ref{defstar}) and ordinary matrix multiplication. The trace
in (\ref{NCYM}) is over the matrix indices of the fields. This writing of
ordinary Yang-Mills theory on the noncommutative space as the noncommutative
gauge theory (\ref{NCYM}) on an ordinary space transmutes the $U(N)$ colour
degrees of freedom into spacetime degrees of freedom along the noncommutative
directions. The former gauge theory can be expressed in terms of Weyl
operators, as we will describe in the next section.

The action (\ref{NCYM}) is invariant under the infinitesimal noncommutative
gauge transformation $A_i\mapsto A_i+\delta_\lambda A_i$, where
$\lambda=\lambda^\dagger\in\mat_N(\alg_\theta)$ and
\bea
\delta_\lambda A_i&=&\del_i \lambda +i\,[\lambda,A_i]^{~}_\star \ , \nn\\
\delta_\lambda F_{ij}&=&i\,[\lambda,F_{ij}]^{~}_\star \ .
\label{ncgaugetransf}\eea
It is straightforward to show that the commutator of two such transformations
with generators $\lambda,\lambda'\in\mat_N(\alg_\theta)$ is given by
\be
[\delta_\lambda,\delta_{\lambda'}]A_i=\delta_{i\,
[\lambda,\lambda']^{~}_\star}A_i \ .
\label{deltacomm}\ee
This implies that the subspace of $\mat_N(\alg_\theta)$ which generates the
noncommutative gauge transforms (\ref{ncgaugetransf}) is a Lie algebra with
respect to the Moyal bracket. In the following we will present an explicit
description of this Lie algebra.

\subsection{Noncommutative Gauge Symmetry}

We will first define more precisely what is meant by a noncommutative gauge
symmetry. The algebra ${\cal A}_\theta$ can be represented faithfully on a
separable Hilbert space ${\cal H}$, whose vectors we interpret as fundamental
matter
fields. The space $\hil$ is an $\alg_\theta$-module which corresponds to a
vector bundle over noncommutative $\real^d$. For example, we can take
$\hil=\hil_{\rm M}$, where
\beq
\hil_{\rm M}=L^2\left(\real^d\,,\,\dd^dx\right)\otimes\complex^N \ ,
\label{hilL2Rd}\eeq
with the (left) action of $\alg_\theta$ defined as
\beq
f\,:\,\psi~\longmapsto~f\star\psi
\label{fstaraction}\eeq
for $f\in\alg_\theta$ and $\psi\in\hil_{\rm M}$. This representation is
reducible, but this is not surprising, because the usual representation of the
commutative algebra of functions $\alg_0$ as operators on Hilbert space is also
a highly reducible representation. Namely, we have
\beq
L^2\left(\real^d\,,\,\dd^dx\right)=\int\limits_{x\in\reals^d}
\!\!\!\!\!\!\!\!\!\!~\!\!\ominus~\delta_x \ ,
\label{commreducible}\eeq
where $\delta_x:\alg_0\to\complex$ is the evaluation functional at
$x\in\real^d$ which is the character of the commutative algebra $\alg_0$ given
by
\beq
\delta_x(f)=\int\dd^dy~\delta(x-y)\,f(y) \ , ~~ f\in\alg_0 \ ,
\label{evalf}\eeq
and which defines a one-dimensional irreducible representation of $\alg_0$ on
$L^2(\real^d,\dd^dx)$ via pointwise multiplication as
\beq
\delta_x(f)\cdot\psi=f(x)\,\psi \ .
\label{chixfpsi}\eeq

In the noncommutative case, the invariant subspaces are larger and the
reducibility of the representation is diminished. To see this explicitly, it is
convenient to exploit global Euclidean invariance and rotate to a basis of
$\real^d$
in which the matrix $\theta_{ij}$ assumes its canonical skew-diagonal form with
skew-eigenvalues $\theta_a$, $a=1,\dots,\frac d2$. Here and in the following
we will assume, for ease of notation, that all $\theta_a$ are positive.
In this basis the coordinate
operators split into $\frac d2$ mutually commuting blocks in each of which the
commutation relation
\beq
[x_{2a-1},x_{2a}]=i\,\theta_a \ , ~~ a=1,\dots,\mbox{$\frac d2$} \ ,
\label{skewcommrel}\eeq
is satisfied. The algebra (\ref{skewcommrel}) can then be represented by the
operators
\beq
x_{2a}=Q_a \ , ~~ x_{2a-1}=i\,\theta_a\,\frac\partial{\partial Q_a}
{}~~~~({\rm no}~{\rm sum}~{\rm on}~a)
\label{Schrodinger}\eeq
acting on the usual quantum mechanical Hilbert space
\beq
\hil_{\rm Q}=\bigotimes_{a=1}^{d/2}L^2(\real,\dd Q_a) \ .
\label{QMhil}\eeq
By the Stone-von~Neumann theorem, the Schr\"odinger representation
(\ref{Schrodinger},\ref{QMhil}) is the unique unitary irreducible
representation of the Heisenberg algebras (\ref{skewcommrel}). The space
$\hil_{\rm Q}\otimes\complex^N$ defines a proper, diagonal subspace of the
Hilbert space (\ref{hilL2Rd}), thus proving reducibility of the representation
(\ref{fstaraction}). An invariant subspace can be easily constructed, for
example, by introducing the complex coordinates
\beq
z_a=\frac1{\sqrt{2\theta_a}}\,\Bigl(x_{2a-1}-i\,x_{2a}\Bigr) \ ,
\label{complexcoord}\eeq
and the Bargmann space $\hil_{\rm B}$ of coherent state wavefunctions
\beq
\psi^{~}_{\rm B}(z,z^*)=F(z)~\e^{-z^\dagger z} \ ,
\label{psiB}\eeq
where $F(z)$ is an $N\times N$ matrix-valued holomorphic function of
$z=(z_a)_{a=1}^{d/2}\in\complex^{d/2}$. It is then straightforward to check
that $\hil_{\rm B}$ is an invariant subspace for the action (\ref{fstaraction})
of
$\alg_\theta$ on the Hilbert space (\ref{hilL2Rd}).

In analogy with the commutative case, we may then consider the gauge
transformations
\be
\psi~\longmapsto~\psi^U=U\star\psi \ ,
\label{Ugaugetransf}\ee
where $\psi\in\hil_{\rm M}$ and $U$ is a unitary element of the algebra
$\mat_N({\cal A}_\theta^+)$. Note that since $\mat_N({\cal A}_\theta)$ is not a
unital algebra, it is necessary to add an identity element to $\alg_\theta$ and
work with the algebra $\alg_\theta^+=\alg_\theta\oplus\complex$ in order to
properly define unitary elements~\cite{Weggeolsen}. Geometrically, this
extension corresponds to considering functions on the one-point
compactification of $\real^d$.\footnote{\baselineskip=12pt Such a
unitalization has been used recently in the construction of noncommutative
instantons~\cite{Schwarz}.} The transformations (\ref{Ugaugetransf})
preserve the representation of $\alg_\theta$ on $\hil_{\rm M}$, and
$\star$-unitarity
\beq
U\star U^\dagger=U^\dagger\star U=\id
\label{starunitarity}\eeq
guarantees that they preserve the Hilbert norm of the matter
field $\psi$. The gauge transformation (\ref{Ugaugetransf}) is a
deformation of that for an ordinary $U(N)$ gauge theory.

The action of the covariant derivative (\ref{gaugecovderiv}) as an operator on
$\hil_{\rm M}$ is defined by
\beq
\nabla_i(\psi)=\partial_i\psi-i\,A_i\star\psi \ .
\label{nablahil}\eeq
Since from the definition (\ref{defstar}) we have
\beq
\partial_i(f\star\psi)=(\partial_if)\star\psi+f\star\partial_i\psi \ ,
\label{Leibnizstar}\eeq
it follows that $\nabla_i$ satisfies a (left) Leibnitz rule with respect to
the representation of $\alg_\theta$ on $\hil_{\rm M}$,
\be
\nabla_i(f\star\psi)=\left(\nabla^{\rm ad}_if\right)\star\psi
+f\star\nabla_i(\psi) \ ,
\label{Leibniznabla}\ee
where the adjoint gauge connection $\nabla_i^{\rm ad}$ is the linear derivation
of the algebra $\alg_\theta$ defined by
\beq
\nabla_i^{\rm ad}f=\partial_if+i\,[f,A_i]^{~}_\star \ .
\label{nablaadjoint}\eeq
{}From (\ref{Leibniznabla}) it follows that $\nabla_i(\psi)$ lies in the same
representation of $\alg_\theta$ as the matter field $\psi$, and so it should
transform in the same way (\ref{Ugaugetransf}) as $\psi$ under gauge
transformations. This requires the covariant derivatives to transform in the
adjoint representation
\beq
\nabla_i~\longmapsto~\nabla_i^U~~~~{\rm with}~~\nabla_i^U(\psi)=U\star
\nabla_i\left(U^\dagger\star\psi\right) \ ,
\label{nablaiU}\eeq
so that $\nabla_i^U(\psi^U)=\nabla_i(\psi)^U$.

We see therefore that gauge transformations in noncommutative Yang-Mills theory
are determined by inner automorphisms $f\mapsto U\star f\star U^\dagger$ of
the algebra $\mat_N(\alg_\theta^+)$. These transformations correspond to
rotations of the algebra elements, and they are parametrized by unitary
elements $U$ of $\mat_N({\cal A}_\theta^+)$ which form the infinite dimensional
group
$U(\mat_N(\alg_\theta^+))$. Normally, when it is said that a gauge group is
finite dimensional, what is really meant is that a gauge transformation is
a map from the spacetime manifold into a finite dimensional Lie group. But
the map itself is an element of an infinite dimensional group,
$U(\mat_N(\alg_0^+))\cong U(\alg_0^+)\otimes U(N)$, with $U(\alg_0^+)$ the
group of $S^1$-valued functions. In the noncommutative case the gauge group is
itself infinite dimensional, and it is non-abelian even in the simplest
instance
of $U(1)$ gauge symmetry. In this paper we will try to understand the Lie
algebraic structure related to this group. Unlike the commutative case,
however, the group $U(\mat_N(\alg_\theta^+))$ is not the tensor product of
a function space with a finite dimensional Lie group, because of the mixing
of internal $U(N)$ and spacetime degrees of freedom that we have alluded
to. This will lead to a much richer algebraic structure.

In the infinite dimensional case, there is no guarantee that we can use
the usual construction of a Lie algebra starting from the tangent space to
the group at the identity~\cite{KrieglMichor}. In fact, it is
known~\cite{murphy} that while elements of the form $\e^{i\,\lambda}$, with
$\lambda$ self-adjoint, are unitary, the converse is not generally true.
However, we can circumvent these problems in the present case by exploiting the
fact that we are really defining the group $U(\mat_N(\alg_\theta^+))$ by its
representation as operators on a Hilbert space, as in (\ref{Ugaugetransf}).
Given an $\alg_\theta$-module ${\cal H}$, let $U\mapsto{\cal U}_{\,U}$ be a
continuous unitary representation of $U(\mat_N(\alg_\theta^+))$ on $\hil$. Then
for each fixed element $U\in U(\mat_N(\alg_\theta^+))$, the operators ${\cal
U}_{\,t\,U}$, $t\in\real$, form a one-parameter transformation group and we may
define an operator on $\hil$ by
\beq
\lambda_U=\lim_{t\to0}\,\frac1{i\,t}\,\Bigl({\cal U}_{\,t\,U}-\id_\hil
\Bigr) \ .
\label{infops}\eeq
By Stone's theorem~\cite{ReedSimon}, the infinitesimal operators (\ref{infops})
are essentially self-adjoint, and we may write ${\cal
U}_{\,t\,U}=\e^{i\,t\,\lambda_U}$. In this way it makes sense to speak of the
Lie algebra of the group $U(\mat_N(\alg_\theta^+))$. Indeed, this is the
situation that is anticipated from the relationships between noncommutative
gauge theories and matrix models.

We will thereby consider the elements of a basis for the Hermitian elements
$u(\mat_N({\cal A}_\theta))$ and study the commutation relations between
them. In this paper we will limit ourselves to the case of a $U(1)$ gauge
symmetry. The gauge algebra for the cases $N>1$ can then be obtained from
the Lie algebraic tensor product between the algebra we will find and the
$u(N)$ Lie algebra. As we will discuss later on, the gauge algebra
$u(\alg_\theta)$ of noncommutative electrodynamics contains all unitary
gauge algebras $u(N)$ in a very precise and exact way, i.e. $U(1)$
noncommutative Yang-Mills theory contains all possible noncommutative gauge
theories with non-abelian unitary gauge groups~\cite{GN,BLP}. This will turn 
out, in fact, to be a very important geometric feature of the noncommutative
gauge group. We will therefore consider $\star$-unitary elements of the form
$U(x)=\e^{\star\,i\,\lambda(x)}$, where $\lambda(x)$ is a real-valued function
on $\real^d$ and the $\star$-exponential is defined by the understanding
that all products in its Taylor series representation are $\star $-products.
The infinitesimal gauge transform of the covariant derivative is then given by
\beq
\nabla_i(\psi)~\longmapsto~\nabla_i(\psi)+i\,\lambda\star\nabla_i(\psi)
-i\,\nabla_i(\lambda\star\psi) \ ,
\label{nablaitransfinf}\eeq
which from (\ref{nablahil}) immediately implies the gauge transformation rule
(\ref{ncgaugetransf}). Generically then, the gauge algebra acts on elements
of the algebra $\alg_\theta$ via the homogeneous, covariant transformations
\be
f~\longmapsto~f+i\,[\lambda,f]^{~}_\star \ .
\label{fhomo}\ee

There is a very important property of the algebra of noncommutative
$\real^d$ that we will exploit in the following. Unlike the case of
ordinary $\real^d$, where the algebra $\alg_0$ of functions is commutative
and there are no inner automorphisms, the gauge symmetries here act via
rotations of functions and correspond to internal fluctuations of the
spacetime geometry in the above sense. The inner automorphisms form a
normal subgroup ${\rm Inn}(\alg_\theta)$ of the automorphism group ${\rm
Aut}(\alg_\theta)$ of the algebra $\alg_\theta$. The exact sequence of groups
\beq
\id~\longrightarrow~{\rm Inn}(\alg_\theta)~\longrightarrow
{}~{\rm Aut}(\alg_\theta)~\longrightarrow~{\rm Out}(\alg_\theta)~
\longrightarrow~\id
\label{exactseq}\eeq
defines the remaining outer automorphisms of $\alg_\theta$ such that the full
automorphism group ${\rm Aut}(\alg_\theta)$ is the semi-direct product of ${\rm
Inn}(\alg_\theta)$ by the natural action of ${\rm Out}(\alg_\theta)$. For
commutative $\real^d$, there are only outer
automorphisms\footnote{\baselineskip=12pt More precisely, the inner
automorphisms generate ordinary gauge transformations.}, and the group ${\rm
Aut}(\alg_0)$ is naturally isomorphic to the group of diffeomorphisms of
$\real^d$~\cite{realncg}. Given a smooth function $\phi:\real^d\to\real^d$,
there is a natural automorphism $\alpha_\phi:\alg_0\to\alg_0$ defined by
\beq
\alpha_\phi(f)=f\circ\phi^{-1} \ , ~~ f\in\alg_0 \ .
\label{outer}\eeq
Like the inner automorphisms in the noncommutative case, the outer
automorphisms (\ref{outer}) can be represented via unitary conjugation when
the algebra $\alg_0$ is represented by operators on the Hilbert space
(\ref{hilL2Rd}). Given a diffeomorphism $\phi$ of $\real^d$, we may define a
unitary operator $U_\phi$ on $\hil_{\rm M}$ by
\beq
U_\phi\,\psi(x)=\left|\frac{\partial\phi}{\partial x}\right|^{1/2}\,\psi\left(
\phi^{-1}\,x\right) \ , ~~ \psi\in\hil_{\rm M} \ .
\label{Uphipsi}\eeq
In this way, we may identify the automorphism group of the algebra $\alg_0$
with the group $U(\hil_{\rm M})$ of unitary endomorphisms of the Hilbert
space $\hil_{\rm M}$ (or more precisely, as we will see in section~4.2, with
the projective subgroup of this unitary group).

In the noncommutative case we will find that some of these outer
automorphisms are deformed into inner automorphisms and thereby generate gauge
transformations. This will follow from the fact that, generally, the
derivation $\del_i$ generates an infinitesimal automorphism of the algebra.
For the algebra (\ref{commx}) it determines the inner automorphism
\eqn{partialinner} of the ring $\complex[[x_1,\dots,x_d]]/{\cal R}_\theta$. In
fact, since the algebra has a trivial center, any linear derivation of
$\alg_\theta$ can be realized as an inner automorphism of
$\complex[[x_1,\dots,x_d]]/{\cal R}_\theta$. This is a property particular to
noncommutative $\real^d$, because in that case the coordinates $x_i$ generate
the algebra of functions. It does not hold, for example, on the noncommutative
torus. This feature will enable us to use the mixing of gauge and spacetime
degrees of freedom to realize certain geometric transformations of $\real^d$ as
genuine gauge symmetries of
noncommutative Yang-Mills theory. In this way we will see in fact that the
automorphism group of the noncommutative algebra $\alg_\theta$ lies in between
the two extremes generated by the commutative algebra $\alg_0$, for which ${\rm
Aut}(\alg_0)={\rm Out}(\alg_0)$, and a finite-dimensional matrix algebra
$\mat_N(\complex)$ for which all automorphisms are inner automorphisms.

\newsection{Explicit Presentations}

In this section we will construct two explicit representations of the Lie
algebra $u(\alg_\theta)$, each of which will be useful in its own right in the
following. The first one will be important for the representation $\hil_{\rm
M}$ of the algebra $\alg_\theta$ in terms of fundamental matter fields, while
the
second one will be pertinent to the Schr\"odinger $\alg_\theta$-module
$\hil_{\rm Q}$ which will be used later on to explicitly identify the
noncommutative gauge algebra.

\subsection{Symmetric Representation}

Using the formulas $x_i^{\star n^i}=x_i^{n^i}$ and $x_i\star
x_j=x_ix_j+\frac i2\,\theta_{ij}$, it is possible to express the
real-valued gauge function $\lambda(x)$ as a series in $\star$-monomials
generated by the $x_i$'s (with the appropriate convergence criterion on the
expansion coefficients). However, although the $x_i$ are Hermitian
operators, the noncommutativity of the coordinates implies in general that
$x_i^{n^i}\star x_j^{n^j}\neq x_j^{n^j}\star x_i^{n^i}$ for $i\neq j$, and
so the $\star$-monomials do not constitute a good basis for the space of
Hermitian operators. We need to use appropriate Hermitian combinations of
the $\star$-monomials in the expansions of the elements of the algebra. We
will choose as basis
\be
T_{\vec n}(x)=\NO x_1^{\star n^1}\star\cdots\star x_d^{\star n^d}\NO \ ,
\label{defT}\ee
where $\vec n=(n^1,\dots,n^d)$ is a $d$-dimensional vector of non-negative
integers, and we have defined the symmetric $\star$-product of functions by
\be
\NO f_1\star\cdots\star f_m\NO=\frac1{m!}\,\sum_{\sigma\in S_m}
f_{\sigma(1)}\star\cdots\star f_{\sigma(m)} \ .
\label{Symstar}\ee
In other words, the operator $T_{\vec n}$ is given by a sum over all
permutations $\sigma\in S_{|\vec n|}$, $|\vec n|\equiv\sum_in^i$, of
$\star$-products of the coordinates such that $x_i$ appears exactly $n^i$ times
in each $\star$-monomial. In the following we will define the
$\star$-symmetrization of a functional ${\cal F}(f_1,\dots,f_m)$ of $m$
functions $f_1,\dots,f_m$ by first formally expanding $\cal F$ as a Taylor
series and then applying the symmetrization operation to each monomial,
\beq
\NO{\cal F}(f_1,\dots,f_m)\NO=\sum_{\vec k\in\zeds_+^m}\frac{
{\cal F}^{(k^1,\dots,k^m)}(0,\dots,0)}{k^1!\cdots k^m!}~\NO f_1^{k^1}
\star\cdots\star f_m^{k^m}\NO \ .
\label{Symfn}\eeq

The operators (\ref{defT}) are Hermitian and constitute an improper basis
of the algebra $u(\alg_\theta)$, in fact they span the ring
$\complex[[x_1,\dots,x_d]]/{\cal R}_\theta$. With the use of the commutation
relations
\eqn{commx}, one can easily see that (\ref{defT}) coincides with the
ordinary, undeformed monomial product. Thus a generic Schwartz function
$\lambda(x)$ on $\real^d$ can be expanded as
\be
\lambda=\sum_{\vec n\in\zeds_+^d}\frac{\partial_1^{n^1}\cdots
\partial_d^{n^d}\lambda(0)}{n^1!\cdots n^d!}~T_{\vec n} \ .
\label{lambdaexp}\ee
This determines the gauge functions of $u(\alg_\theta)$ in terms of a basis
of the vector space spanned by homogeneous symmetric polynomials in the
generators of the Lie algebra generated by the $x_i$'s. In other words,
elements of $u(\alg_\theta)$ lie in the enveloping algebra of the Lie
algebra (\ref{commx}). In the following we will compute the Moyal brackets
of the elements (\ref{defT}),
\be
\Bigl[T_{\vec n}\,,\,T_{\vec m}\Bigr]^{~}_\star =\sum_{\vec p\in\zeds_+^d}
c_{\vec n\vec m\vec p}~T_{\vec p} \label{commt} \ ,
\ee
where $c_{\vec n\vec m\vec p}$ are the structure constants of the Lie
algebra $u(\alg_\theta)$ in the basis (\ref{defT}).

The calculation of the $\star$-commutators \eqn{commt} is most efficiently done
by exploiting the one-to-one correspondence between non-local, noncommutative
fields on a commutative space and local, commutative fields on a
noncommutative space. This is achieved via the Weyl quantization
map~\cite{Weyl}, which associates to every function $f(x)$ on
$\real^d\to\complex$ an operator-valued function of self-adjoint operators
$\hat x_i$ which generate the algebra $[\hat x_i,\hat x_j]=i\,\theta_{ij}$,
and which act faithfully on a particular Hilbert space $\cal H$. Elements
of the algebra $\alg_\theta$ are then represented as operators on $\hil$,
i.e. as elements of the endomorphism algebra ${\rm End}(\hil)$ of the
Hilbert space. We will denote this representation by $\alg_\theta(\hil)$.
The Weyl map $\Omega:\alg_0\to\alg_\theta({\cal H})$ is given by
\be
F(\hat x)=\Omega(f)=\int\dd^dx~f(x)\int\frac{\dd^d\xi}{(2\pi)^d}~
\e^{i\,\xi{\cdot}(\hat x-x)} \ ,
\label{Weyl}\ee
and it generalizes the transformation which is usually used in quantum
mechanics to associate a quantum operator to a function on a classical
phase space. The Weyl transform has an inverse, known as the Wigner map
\cite{Wigner}, which is given by
\be
f(x)=\Omega^{-1}(F)=\pi^{d/2}~{\rm Pfaff}(\theta)\,
\int\frac{\dd^d\xi}{(2\pi)^d}~\Tr_\hil\Bigl(F(\hat x)\,
\e^{i\,\xi{\cdot}(\hat x-x)}\Bigr) \ ,
\label{Wigner}\ee
where the trace over states of $\hil$ is equivalent to integration over
the noncommuting coordinates $\hat x_i$. The useful property of the Weyl and
Wigner maps is that they generate an isomorphism
$\alg_\theta\leftrightarrow\alg_\theta(\hil)$, i.e.
\bea
\Omega(f\star g)&=&\Omega(f)\,\Omega(g) \ , \nn\\
\Omega^{-1}(F\,G)&=&\Omega^{-1}(F)\star \Omega^{-1}(G) \ .
\label{WWcorr}\eea

Moreover, spacetime averages of fields map to traces of Weyl operators,
\beq
\int\dd^dx~f(x)=\pi^{d/2}~{\rm Pfaff}(\theta)~\Tr_\hil
\Bigl(\Omega(f)\Bigr) \ ,
\label{Trint}\eeq
and under the Weyl-Wigner correspondence the $U(1)$ noncommutative gauge
theory (\ref{NCYM}) becomes ordinary Yang-Mills theory on the noncommutative
space,
\beq
S=\frac{\pi^{d/2}~{\rm Pfaff}(\theta)}4\,
\Tr_\hil\Bigl(\Omega(F_{ij})\,\Omega(F^{ij})\Bigr) \ .
\label{WWNCYM}\eeq
In this way we may regard noncommutative gauge theory as ordinary
Yang-Mills theory with the extended, infinite dimensional local gauge
symmetry algebra $u(\alg_\theta)$. Note that from (\ref{WWcorr}) it follows
directly that the $\star$-product possesses the same algebraic properties as
the ordinary operator product in ${\rm End}(\hil)$, i.e. it is associative but
noncommutative, while (\ref{Trint}) shows explicitly that spacetime integrals
of $\star$-products of functions have precisely the same cyclic permutation
symmetries as traces of operator products (c.f.~(\ref{starint})).

In the present case, the real advantage of using the Weyl-Wigner correspondence
is the operator ordering that is provided by the map (\ref{Weyl}). By
definition, it symmetrically orders operator products. This implies that if we
define the Weyl operators
\be
\hat T_{\vec n}\equiv\Omega\left(x_1^{n^1}\cdots  x_d^{n^d}\right) \ ,
\ee
then $\hat T_{\vec n}=\Omega(T_{\vec n})$ and the Lie algebra \eqn{commt} can
be computed as
\be
\left[\hat T_{\vec n}\,,\,\hat T_{\vec m}\right]=\sum_{\vec p\in\zeds_+^d}
c_{\vec n\vec m\vec p}~\hat T_{\vec p} \ .
\label{hatTcomm}\ee
A further simplification comes from rotating to a basis in which the
commutation relations of the noncommutative space assume the skew-block form
(\ref{skewcommrel}). By denoting $\hat T_{\vec
n}^{(a)}=\Omega(x_{2a-1}^{n^{2a-1}}x_{2a}^{n^{2a}})$, the left-hand side of
(\ref{hatTcomm}) may then be computed from
\be
\left[\hat T_{\vec n}\,,\,\hat T_{\vec m}\right]=\sum_{a=1}^{d/2}\hat T_{\vec n
+\vec m}^{(1)}\cdots \hat T_{\vec n+\vec m}^{(a-1)}
\left[\hat T_{\vec n}^{(a)}\,,\,\hat T_{\vec m}^{(a)}\right]
\hat T_{\vec n+\vec m}^{(a+1)}\cdots \hat T_{\vec n+\vec m}^{(d)} \ ,
\label{hatTcommsplit}\ee
and the commutators appearing on the right-hand side of (\ref{hatTcommsplit})
can be calculated as
\be
\left[\hat T_{\vec n}^{(a)}\,,\,\hat T_{\vec m}^{(a)}\right]=\Omega\Bigl(
(x_{2a-1}^{n^{2a-1}}x_{2a}^{n^{2a}})\star(x_{2a-1}^{m^{2a-1}}x_{2a}^{m^{2a}})
-(x_{2a-1}^{m^{2a-1}}x_{2a}^{m^{2a}})\star(x_{2a-1}^{n^{2a-1}}x_{2a}^{n^{2a}})
\Bigr) \ .
\label{moncomm}\ee
We are therefore left with a simple calculation which involves only the
$\star$-products of monomials, and not the combinatorics involved in the
symmetrization operation.

The computation of (\ref{moncomm}) is straightforward. For this, we note that
in a given skew-block $a$, the $\star$-product (\ref{defstar}) may be written
as
\bea
(f\star g)^{(a)}&=&\sum_{r=0}^\infty~\sum_{s=0}^\infty\frac{(-1)^r\,(i\,
\theta_a)^{r+s}}{2^{r+s}\,r!\,s!}\,\frac{\partial^{r+s}f}{\partial x_{2a-1}^r
\partial x_{2a}^s}\,\frac{\partial^{r+s}g}{\partial x_{2a-1}^s\partial
x_{2a}^r}\nn\\&=&\sum_{n=0}^\infty\left(\frac{i\,\theta_a}2\right)^n
\,\sum_{r=0}^n\frac{(-1)^r}{(n-r)!\,r!}\,
\left(\partial_{2a-1}^r\partial_{2a}^{n-r}f\right)\left(
\partial_{2a-1}^{n-r}\partial_{2a}^rg\right) \ ,
\label{fstarga}\eea
from which we may express the Moyal bracket (\ref{starcomm}) as
\beq
[f,g]_\star^{(a)}=2i\,\sum_{p=0}^\infty\frac{(\theta_a)^{2p+1}}{4^p}\,
\sum_{l=0}^{2p+1}\frac{(-1)^{p+l}}{l!\,(2p+1-l)!}\,\left(
\partial_{2a-1}^l\partial_{2a}^{2p+1-l}f\right)\left(
\partial_{2a-1}^{2p+1-l}\partial_{2a}^lg\right) \ .
\label{Moyala}\eeq
Setting $f=x_{2a-1}^{n^{2a-1}}x_{2a}^{n^{2a}}$ and
$g=x_{2a-1}^{m^{2a-1}}x_{2a}^{m^{2a}}$ in (\ref{Moyala}) thereby leads to
\bea
&&\left[T_{\vec n}^{(a)}\,,\,T_{\vec m}^{(a)}\right]^{~}_\star=
2i\,\sum_{0\leq p\leq\frac12\bigl(\min(n^{2a-1},m^{2a})+\min(n^{2a},m^{2a-1})
-1\bigr)}\,\frac{(\theta_a)^{2p+1}}{4^p}\nn\\&&~~\times\,\sum_{l=2p+1-
\min(n^{2a},m^{2a-1},2p+1)}^{\min(n^{2a-1},m^{2a},2p+1)}\,\frac{(-1)^{p+l}}
{l!\,(2p+1-l)!}\,\frac{n^{2a-1}!\,n^{2a}!}{(n^{2a-1}-l)!\,(n^{2a}-2p-1+l)!}
\nn\\&&~~\times\,\frac{m^{2a-1}!\,m^{2a}!}{(m^{2a-1}-2p-1+l)!\,(m^{2a}-l)!}
{}~T^{(a)}_{\vec n+\vec m-\underline{(2p+1)}} \ ,
\label{commrs}\eea
where generally $\underline{r}=(r,r,\dots,r)$ denotes the integer vector whose
components are all equal to $r\in\zed_+$.

We can now rotate back to general form by using (\ref{hatTcommsplit}), and
write a commutation relation of the form (\ref{commt}). For the structure
constants of the full Lie algebra we then arrive at
\bea
c_{\vec n\vec m\bigl(\vec n+\vec m-(2p+1)(\vec e_i+\vec e_j)\,\bigr)}&=&
2i\,\frac{(\theta_{ij})^{2p+1}}{4^p}\,\sum_{l=2p+1-
\min(n^j,m^i,2p+1)}^{\min(n^i,m^j,2p+1)}\,\frac{(-1)^{p+l}}
{l!\,(2p+1-l)!}\nn\\&&\times\,\frac{n^i!\,n^j!}{(n^i-l)!\,(n^j-2p-1+l)!}
\,\frac{m^i!\,m^j!}{(m^i-2p-1+l)!\,(m^j-l)!}\nn\\&& { ~~~ }\nn\\
c_{\vec n\vec m\vec k}=0
{}~~~~{\rm for}~~\vec k&\neq&\vec n+\vec m-(2p+1)\left(\vec e_i+\vec e_j
\right) \ , \non
2p+1&>&\max_{i<j}\Bigl\{\min(n^i,m^j)+\min(n^j,m^i)\Bigr\} \ ,
\label{cnmp}\eea
where $\vec e_i$ is the standard basis of the hypercubic lattice $\zed_+^d$,
\beq
\left(\vec e_i\right)^j=\delta_i^j \ .
\label{vecei}\eeq
Although the expressions (\ref{commrs}) and (\ref{cnmp}) as they stand are not
particularly transparent, we will see in the following how their explicit
expansions in $\theta_{ij}$ reveal some remarkable geometrical features of the
gauge algebra $u(\alg_\theta)$.

\subsection{Density Matrix Representation}

Another very important representation of the Lie algebra $u(\alg_\theta)$
arises on the space of Weyl operators $\alg_\theta(\hil)$ when $\hil$ is taken
to be the Hilbert space (\ref{QMhil}) of quantum mechanics (with respect to the
skew-diagonalization of $\theta_{ij}$), i.e. the Schr\"odinger
representation of the Heisenberg commutation relations (\ref{skewcommrel}). Let
us fix, as before, an integer vector $\vec n\in\zed_+^d$, and for each
$a=1,\dots,\frac d2$ consider the Weyl operators $\hat z_a$ corresponding to
the complex coordinates (\ref{complexcoord}). These operators obey the
commutation relations
\beq
\left[\hat z_a\,,\,\hat z_a^\dagger\right]=1
\label{hatzacommrel}\eeq
in each commuting skew-block, and the Weyl transform (\ref{Weyl}) may be
expressed in terms of them via the substitutions
\beq
\hat x_{2a-1}=\sqrt{\frac{\theta_a}2}\,\left(\hat z_a+\hat z_a^\dagger\right)
\ , ~~ \hat x_{2a}=i\,\sqrt{\frac{\theta_a}2}\,\left(\hat z_a-
\hat z_a^\dagger\right) \ .
\label{hatxhatzrel}\eeq
The Hilbert space (\ref{QMhil}) may then be represented in terms of the
standard Fock space of creation and annihilation operators as
\beq
L^2(\real,\dd Q_a)=\ell^2(\zed_+)=
\bigoplus_{n^{2a}=0}^\infty\complex|n^{2a}\rangle \ ,
\label{Fock}\eeq
where $|n^{2a}\rangle$ are the orthonormal eigenstates of the number operator
$\hat z_a^\dagger\,\hat z_a$ with eigenvalues $n^{2a}\in\zed_+$, and the action
of the Weyl operators is defined by
\beq
\hat z_a|n^{2a}\rangle=\sqrt{n^{2a}}\,
|n^{2a}-1\rangle\ , ~~ \hat z_a^\dagger|n^{2a}\rangle=
\sqrt{n^{2a}+1}\,|n^{2a}+1\rangle \ .
\label{Fockstates}\eeq

A basis for the Lie algebra $u(\alg_\theta(\hil_{\rm Q}))$ of Hermitian Weyl
operators on the Hilbert space (\ref{QMhil}) is given by the
basis for density matrices,
\beq
\hat\Sigma_{\vec n}^{(\epsilon_1\cdots\epsilon_{d/2})}=\bigotimes_{a=1}^{d/2}
\hat\Sigma_{\vec n}^{(a)\epsilon_a} \ , ~~ \epsilon_a=\pm \ ,
\label{Pihatdef}\eeq
where
\bea
\hat\Sigma^{(a)+}_{\vec n}&=&i\,|n^{2a}\rangle\langle
n^{2a-1}|-i\,|n^{2a-1}\rangle\langle n^{2a}| \ ,
\non\hat\Sigma^{(a)-}_{\vec n}&=&|n^{2a}\rangle\langle
n^{2a-1}|+|n^{2a-1}\rangle\langle n^{2a}|
\label{taualphahatdef}\eeq
for each $a=1,\dots,\frac d2$ span the space of self-adjoint operators on the
Fock space (\ref{Fock}). To compute the Wigner functions (\ref{Wigner})
corresponding to the Weyl operators (\ref{Pihatdef}), we use the standard
Groenewold distribution functions for the energy eigenstates of the
one-dimensional harmonic oscillator~\cite{Groenewold,CUZ}
\bea
E_{n^{2a-1}n^{2a}}(z_a,z_a^*)&\equiv&
\Omega^{-1}\left(\,|n^{2a-1}\rangle\langle n^{2a}|\,
\right)\non&=&\frac1{2\pi}\,\int\limits_{-\infty}^\infty\dd\xi_{2a}~
\e^{-i\,\xi_{2a}x_{2a}}\,\langle x_{2a-1}-
\mbox{$\frac{\theta_a}2$}\,\xi_{2a}|n^{2a-1}
\rangle\langle n^{2a}|x_{2a-1}+\mbox{$\frac{\theta_a}2$}\,\xi_{2a}
\rangle\non&=&\frac{(-1)^{n^{2a}}}\pi\,\sqrt{\frac{n^{2a}!}{n^{2a-1}!}}
\,\left(4\theta_a|z_a|^2\right)^{\frac{n^{2a-1}-n^{2a}}2}~\e^{-2\theta_a
|z_a|^2}~\e^{i\,(n^{2a-1}-n^{2a})\,{\rm arg}(z_a)}\non&&\times\,
L_{n^{2a}}^{n^{2a-1}-n^{2a}}\left(4\theta_a|z_a|^2\right) \ ,
\label{HOWigner}\eea
where $z_a$ are the complex coordinates (\ref{complexcoord}) and
\beq
L_n^\beta(t)=\frac{t^{-\beta}~\e^t}{n!}\,\frac{d^n}{dt^n}\,
\Bigl(t^{\beta+n}~\e^{-t}\Bigr)
\label{Laguerre}\eeq
are the associated Laguerre functions. Since
$E_{n^{2a-1}n^{2a}}=E_{n^{2a}n^{2a-1}}^*$, the Wigner functions
which generate the gauge algebra $u(\alg_\theta)$ in the occupation number
basis of the Schr\"odinger representation are thereby given as
\bea
\Sigma_{\vec n}^{(\epsilon_1\cdots\epsilon_{d/2})}(x)&=&\prod_{a=1}^{d/2}
\Sigma_{\vec n}^{(a)\epsilon_a}(z_a,z_a^*) \ , ~~ \epsilon_a=\pm \ ,
\non
\Sigma_{\vec n}^{(a)+}(z_a,z_a^*)&=&-\frac{2(-1)^{n^{2a}}}\pi\,
\sqrt{\frac{n^{2a}!}{n^{2a-1}!}}
\,\left(4\theta_a|z_a|^2\right)^{\frac{n^{2a-1}-n^{2a}}2}~\e^{-2\theta_a
|z_a|^2}\non&&\times\,\sin\Bigl((n^{2a-1}-n^{2a})\,{\rm arg}(z_a)\Bigr)\,
L_{n^{2a}}^{n^{2a-1}-n^{2a}}\left(4\theta_a|z_a|^2\right) \ , \non
\Sigma_{\vec n}^{(a)-}(z_a,z_a^*)&=&\frac{2(-1)^{n^{2a}}}\pi\,
\sqrt{\frac{n^{2a}!}{n^{2a-1}!}}
\,\left(4\theta_a|z_a|^2\right)^{\frac{n^{2a-1}-n^{2a}}2}~\e^{-2\theta_a
|z_a|^2}\non&&\times\,\cos\Bigl((n^{2a-1}-n^{2a})\,{\rm arg}(z_a)\Bigr)\,
L_{n^{2a}}^{n^{2a-1}-n^{2a}}\left(4\theta_a|z_a|^2\right) \ .
\label{Wignernumber}\eea

The commutation relations of the operators (\ref{taualphahatdef}) can be easily
worked out using the orthonormality relations $\langle m|n\rangle=\delta_{nm}$,
and for the Moyal brackets of the functions (\ref{Wignernumber}) we thus find
\bea
\left[\Sigma^{(a)+}_{\vec n}\,,\,\Sigma^{(a)+}_{\vec m}\right]^{~}_\star&=&-i
\left(\delta_{n^{2a-1}m^{2a-1}}\,\Sigma^{(a)+}_{n^{2a}m^{2a}}
+\delta_{n^{2a}m^{2a}}\,\Sigma^{(a)+}_{n^{2a-1}m^{2a-1}}\right.\non&&
-\left.\delta_{n^{2a-1}m^{2a}}\,\Sigma^{(a)+}_{n^{2a}m^{2a-1}}
-\delta_{n^{2a}m^{2a-1}}\,\Sigma^{(a)+}_{n^{2a-1}m^{2a}}\right) \ , \non
\left[\Sigma^{(a)+}_{\vec n}\,,\,\Sigma^{(a)-}_{\vec m}\right]^{~}_\star&=&
i\left(\delta_{n^{2a-1}m^{2a-1}}\,\Sigma^{(a)-}_{n^{2a}m^{2a}}-
\delta_{n^{2a}m^{2a}}\,\Sigma^{(a)-}_{n^{2a-1}m^{2a-1}}\right.\non&&
+\left.\delta_{n^{2a-1}m^{2a}}\,\Sigma^{(a)-}_{n^{2a}m^{2a-1}}
-\delta_{n^{2a}m^{2a-1}}\,\Sigma^{(a)-}_{n^{2a-1}m^{2a}}\right) \ , \non
\left[\Sigma^{(a)-}_{\vec n}\,,\,\Sigma^{(a)-}_{\vec m}\right]^{~}_\star&=&
-i\left(\delta_{n^{2a-1}m^{2a-1}}\,\Sigma^{(a)+}_{n^{2a}m^{2a}}
+\delta_{n^{2a}m^{2a}}\,\Sigma^{(a)+}_{n^{2a-1}m^{2a-1}}\right.\non&&
+\left.\delta_{n^{2a-1}m^{2a}}\,\Sigma^{(a)+}_{n^{2a}m^{2a-1}}+
\delta_{n^{2a}m^{2a-1}}\,\Sigma^{(a)+}_{n^{2a-1}m^{2a}}\right) \ .
\label{hattaualphaalg}\eea
It is intriguing to note that the gauge algebra in this representation also
canonically has the structure of a Lie superalgebra. We can define a
$\zed_2$-grading on $u(\alg_\theta)$ by
\beq
{\rm deg}\,\Sigma_{\vec n}^{(a)\epsilon_a}=\frac{1-\epsilon_a}2 \ ,
\label{degdef}\eeq
and compute the $\star$-anticommutators of the odd functions
$\Sigma_{\vec n}^{(a)-}$ to get
\bea
\left\{\Sigma^{(a)-}_{\vec n}\,,\,\Sigma^{(a)-}_{\vec m}\right\}_\star&=&
\delta_{n^{2a-1}m^{2a-1}}\,\Sigma^{(a)-}_{n^{2a}m^{2a}}
+\delta_{n^{2a}m^{2a}}\,\Sigma^{(a)-}_{n^{2a-1}m^{2a-1}}\non&&
+\,\delta_{n^{2a-1}m^{2a}}\,\Sigma^{(a)-}_{n^{2a}m^{2a-1}}+
\delta_{n^{2a}m^{2a-1}}\,\Sigma^{(a)-}_{n^{2a-1}m^{2a}} \ ,
\label{Picommgraded}\eea
where
\beq
\{f,g\}^{~}_\star(x)
=(f\star g)(x)+(g\star f)(x)=\left.2\cos\left(\mbox{$\frac12$}
\,\theta_{ij}\,\partial_{\xi_i}\partial_{\eta_j}\right)\,
f(\xi)g(\eta)\right|_{\xi=\eta=x} \ .
\label{staranticomm}\eeq
It would be interesting to interpret this structure as some sort of ``hidden''
supersymmetry of noncommutative Yang-Mills theory. In any case, with the
definition (\ref{degdef}), the Lie algebra (\ref{hattaualphaalg}) naturally
possesses a multiplicative $\zed_2$-grading.

\subsection{Relations Between the Presentations \label{relations}}

The advantage of the occupation number basis over the basis of
$\star$-monomials is the simplicity of the structure constants in
(\ref{hattaualphaalg}). On the other hand, the Moyal brackets of the symmetric
representation can be written down succinctly in an arbitrary choice of axes
for $\real^d$, contrary to the representation of the previous subsection.
Notice also that the dependence on the noncommutativity parameters in the
density matrix case is completely absorbed into the functions
(\ref{Wignernumber}). This feature makes it difficult to analyse the algebra as
a deformation of that for an ordinary gauge theory. In particular, the
functions do not go over smoothly into a basis of functions for $\alg_0$ in the
commutative limit $\theta_{ij}\to0$. In contrast, the structure constants
(\ref{cnmp}) are amenable to explicit
analysis order by order in the deformation parameters. There are some
interesting relationships between the two bases we have constructed that we
shall now proceed to analyse.

Let us first explicitly describe, for the sake of completeness, the
transformation between the two presentations of the gauge algebra
$u(\alg_\theta)$ given in this section, i.e. the change of basis between the
two sets of functions (\ref{defT}) and (\ref{Wignernumber}). It suffices to do
this in each commuting skew-block $a$. Thus we want to express the functions
$\Sigma^{(a)\pm}_{\vec n}=\Omega^{-1}(\hat\Sigma^{(a)\pm}_{\vec n})$ in terms
of $T_{\vec n}^{(a)}$. Again it is useful to notice that $\Omega^{-1}(\hat
T_{\vec n}^{(a)})=x_{2a-1}^{n^{2a-1}}x_{2a}^{n^{2a}}$. Furthermore, from
(\ref{taualphahatdef}) it follows that
$\Sigma^{(a)\pm}_{n^{2a-1}n^{2a}}=\mp\,\Sigma^{(a)\pm}_{n^{2a}n^{2a-1}}$. Owing
to this latter property we can always assume that $n^{2a-1}\geq n^{2a}$. The
desired change of basis is then given simply by the Taylor series expansions of
the analytic functions $\Sigma_{\vec n}^{(a)\pm}(x)$ in (\ref{Wignernumber}).

For this, we first rewrite the energy distribution functions \eqn{HOWigner} as
\bea
E_{n^{2a-1}n^{2a}}&=&\frac{(-1)^{n^{2a}}}\pi\,\sqrt{\frac{n^{2a}!}{n^{2a-1}!}}
\,\left(4\theta_a\right)^{\frac{n^{2a-1}-n^{2a}}2}\,(x_{2a-1}+i\,
x_{2a})^{n^{2a-1}-n^{2a}}~\e^{-2\theta_a(x_{2a-1}^2+x_{2a}^2)}\non&&\times\,
L_{n^{2a}}^{n^{2a-1}-n^{2a}}\left(4\theta_a(x_{2a-1}^2+x_{2a}^2)\right) \ .
\label{Erewrite}\eea
This expression can be expanded in a Taylor series by using the identity
\cite{Gradshteyn}
\be
L_n^\beta(t+s)=\e^s\,\sum_{m=0}^\infty\frac{(-1)^m}{m!}\,s^m\,L_n^{\beta+m}(t)
\ee
and the explicit expression for the associated Laguerre polynomials
\be
L_n^k(t)=\sum_{m=0}^n(-1)^m\,
\left(\begin{array}{c}n+k\\n-m\end{array}\right)\,\frac{t^m}{m!} \ .
\ee
By using in addition the binomial theorem and the Taylor series expansion of
the exponential function we can expand (\ref{Erewrite}) as
\bea
E_{n^{2a-1}n^{2a}}&=&\frac{(-1)^{n^{2a}}}\pi\,\sqrt{\frac{n^{2a}!}{n^{2a-1}!}}
\,\left(4\theta_a\right)^{\frac{n^{2a-1}-n^{2a}}2}\,
\sum_{k^{2a-1},k^{2a}=0}^\infty\frac{(-2\theta_a\,x_{2a-1}^2)^{k^{2a-1}}}
{k^{2a-1}!}\,\frac{(2\theta_a\,x_{2a}^2)^{k^{2a}}}{k^{2a}!}\non&&\times\,
\sum_{k=0}^\infty\frac{(-4\theta_a\,x_{2a}^2)^k}{k!}~\sum_{l=0}^{n^{2a}}
\left(\begin{array}{c}n^{2a-1}+k\\n^{2a}-l\end{array}\right)\,
\frac{(-4\theta_a\,x_{2a-1}^2)^l}{l!}\non&&\times\,\sum_{p=0}^{n^{2a-1}-n^{2a}}
i^p\,\left(\begin{array}{c}n^{2a-1}-n^{2a}\\p\end{array}\right)\,
x_{2a-1}^{n^{2a-1}-n^{2a}-p}\,x_{2a}^p \ .
\label{Ereexp}\eea
Collecting the powers of $x_{2a-1}$ and $x_{2a}$, and taking the imaginary and
real parts of (\ref{Ereexp}) leads finally to the change of basis from the
$\star$-monomials $T_{\vec n}$ to the Wigner functions $\Sigma_{\vec
n}^{(\epsilon_1\cdots\epsilon_{d/2})}$ of the density matrices,
\bea
\Sigma^{(a)\epsilon_a}_{\vec n}&=&-\frac{2^{\frac{n^{2a-1}-n^{2a}}2+1}\,
\epsilon_a}\pi\,\sqrt{\frac{n^{2a}!}{n^{2a-1}!}}~{\sum_{\vec m}}^{\,\prime}~
\sum_{k=0}^\infty~\sum_{l=0}^{n^{2a}}~\sum_{p=0}^{\bigl[
\frac{n^{2a-1}-n^{2a}-\frac{1-\epsilon_a}2}2\bigr]}(-1)^{
\frac{m^{2a-1}-n^{2a-1}+2p+\frac{1+\epsilon_a}2}2}\non
&&\times\,\frac{2^{k+l}\,\left(\,\sqrt{2\theta_a}\,\right)^{m^{2a-1}+m^{2a}}}
{k!\,l!\,\left(\frac{m^{2a-1}+n^{2a}-n^{2a-1}+2p+\frac{1+\epsilon_a}2}2-l
\right)!\,\left(\frac{m^{2a}-2p-\frac{1+\epsilon_a}2}2-k\right)!}\non
&&\times\,\left(\begin{array}{c}n^{2a-1}+k\\n^{2a}-l\end{array}\right)
\,\left(\begin{array}{c}n^{2a-1}-n^{2a}\\2p+\frac{1+\epsilon_a}2\end{array}
\right)~T_{\vec m}^{(a)} \ .
\label{explbasischange}\eea
The prime on the first sum in (\ref{explbasischange}) means to restrict to
those integer vectors $\vec m$ in each skew-block $a$ for which the parity of
$m^{2a-1}$ is equal to that of $n^{2a-1}-n^{2a}+\frac{1+\epsilon_a}2$ and
the parity of $m^{2a}$ is $\frac{1+\epsilon_a}2$, and that terms involving
factorials of negative integers are to be omitted from the sum. Note in
particular that this basis change shows explicitly how the canonical
$\zed_2$-grading of the occupation number basis is induced by the even/odd
integer grading of the basis $T_{\vec n}$.

As mentioned at the beginning of this section, the two bases $T_{\vec n}$
and $\Sigma_{\vec n}^{(\epsilon_1\cdots\epsilon_{d/2})}$ are well suited
for the representations of the algebra $\alg_\theta$ on the Hilbert spaces
${\cal H}_{\rm M}$ and ${\cal H}_{\rm Q}$, respectively. We will now describe
the relationship between these two spaces as well. For this, we shall exploit
the Weyl-Wigner correspondence and identify the algebra of functions on
noncommutative $\real^d$ with the algebra $\alg_\theta(\hil)$ of Weyl
operators on a certain Hilbert space $\hil$. The group ${\rm
Inn}(\alg_\theta)$ of inner automorphisms (gauge transformations) of the
algebra $\alg_\theta$ is then most efficiently computed via its lift to
this Hilbert space as~\cite{realncg}
\beq {\rm Inn}_\hil(\alg_\theta)=\left\{U\in U(\hil)~\left|~UJ=JU\,,\,
\imath_U\in{\rm Inn}\Bigl(\alg_\theta(\hil)\Bigr)\right.\right\} \ ,
\label{Innhil}\eeq
where $U(\hil)$ is the group of unitary endomorphisms of $\hil$,
\beq
\imath_U(F)=U\,F\,U^{-1} \ , ~~ F\in\alg_\theta(\hil) \ ,
\label{imathdef}\eeq
and $J$ is the Tomita involution which is defined as the anti-linear,
self-adjoint unitary isometry of the Hilbert space $\hil$ such that
$J\alg_\theta(\hil)J^{-1}=\alg_\theta'(\hil)$ is the commutant of the
algebra $\alg_\theta$ in $\hil$. If $\alg_\theta$ acts on $\hil$ from the left
(resp. right), then $J\,\hil$ is a right (resp. left) $\alg_\theta$-module.
The projection $\pi:{\rm Inn}_\hil(\alg_\theta)\to{\rm Inn}(\alg_\theta)$ is
given in terms of the Wigner transform as
\beq
\pi(U)=\Omega^{-1}\,\imath_U \ .
\label{piU}\eeq
In the case that $\hil=\hil_{\rm M}$ is the $L^2$-completion (\ref{hilL2Rd}) of
the algebra $\alg_0$, the canonical involution $J$ is just (complex)
conjugation,
\beq
J(\psi)=\psi^\dagger \ , ~~ \forall\psi\in\hil_{\rm M} \ .
\label{Jpsi}\eeq
In this sense $J$ may be thought of as a charge conjugation operator. When
$\hil=\hil_{\rm Q}$ is the Hilbert space (\ref{QMhil}), i.e. the space of
functions of the complex variables $z_a$, the algebra $\alg_\theta(\hil_{\rm
Q})$ consists of functionals of the operators $z_a$ and $\frac\partial{\partial
z_a}$, with the representation $z_a\mapsto z_a$ and
$z_a^*\mapsto\frac\partial{\partial z_a}$ on $\hil_{\rm Q}$. The symmetry $J$
then effectively enlarges the Hilbert space $\hil_{\rm Q}$ to $\hil_{\rm M}$,
with  $\alg_\theta'(\hil_{\rm Q})$ the space of functionals of the operators
$z_a^*$ and $\frac\del{\del z_a^*}$. This algebra is naturally isomorphic to
$\alg_\theta(\hil_{\rm Q})$.

The bi-module structure induced by the operator $J$ in (\ref{Innhil}) can be
motivated physically within the context of open string quantization in
background Neveu-Schwarz fields. Quantizing the point particle at a given
endpoint of an open string produces a Hilbert space $\hil$. In the
Seiberg-Witten scaling limit $\alpha'\,\theta_i^{~j}\to\infty$~\cite{SW},
whereby the string oscillations can be neglected, the imposition of identical
boundary conditions at both endpoints of an open string yields a Hilbert space
of the form $\hil\otimes\hil^\vee$, where $\hil^\vee$ is the complex conjugate
$\alg_\theta$-module to $\hil$ corresponding to the opposite orientations of a
pair of string endpoints. We may naturally identify the Hilbert spaces
$\hil=\hil_{\rm Q}$~\cite{SW} and $\hil^\vee=J\,\hil$. The property in
(\ref{Innhil}) then reflects the fact that a lifted gauge transformation should
preserve the actions of $\alg_\theta$ at opposite ends of the open strings. We
can therefore interpret the density matrix representation of the gauge algebra
$u(\alg_\theta)$ as that which pertains to a single endpoint of an open string,
i.e. the unoriented case of Type I superstrings, while the $\star$-monomial
representation is related to Type II superstrings and the presence of both
endpoints corresponding to a chosen relative orientation. In this way the
quotient by the real structure $J$ maps Type II D-branes onto Type I D-branes
and their associated orientifold planes. This has been used in~\cite{Jgauge} to
construct noncommutative gauge algebras based on non-unitary groups.

\newsection{Noncommutative Canonical Transformations}

In this section we will explore the geometric spacetime transformations which
are induced by noncommutative gauge transformations. These can all be
interpreted in terms of the symplectic geometry induced on $\real^d$ by the
constant antisymmetric tensor $\theta_{ij}$. We will see in fact that the gauge
algebra of noncommutative Yang-Mills theory is a deformation of the algebra of
symplectomorphisms of $\real^d$ in a very precise and analytical way.

\subsection{Quantum Deformation of the Poisson Algebra}

As we have discussed in section~2.2, the inner automorphisms of the algebra
$\alg_\theta$ can be regarded in noncommutative geometry as the counterpart
of ``point transformations'' in a space in which it is not really appropriate
to speak of points. Infinitesimal inner automorphisms of the form (\ref{fhomo})
are necessarily volume preserving, since the corresponding spacetime
averages transform as
\be
\int\dd^dx~f(x)~\longmapsto~\int\dd^dx~f(x)+i\,\int\dd^dx~[\lambda,f]^{~}_\star
(x) \ .
\label{avgtransf}\ee
Because of (\ref{starint}), the integral of the $\star$-commutator in
(\ref{avgtransf}) vanishes, and so the volume integral of any covariant
function is invariant under inner automorphisms. Thus, the volume preserving
diffeomorphisms have a very special place in noncommutative gauge theory. They
are defined infinitesimally by the transformations $f\mapsto f+\delta_Vf$ of
functions $f\in\alg_0$ by
\be
\delta_Vf(x)=V(f)(x)=V^i(x)\,\partial_if(x)~~~~~~{\rm with}~~\partial_i
V^i(x)=0 \ .
\label{deltaVf}\ee
Via an integration by parts we may deduce from the divergence-free
condition of (\ref{deltaVf}) that $\int\dd^dx~\delta_Vf(x)=0$. Therefore,
noncommutative gauge symmetries {\it cannot} realize arbitrary
diffeomorphims, but rather only the subalgebra of volume-preserving
transformations of the spacetime. The natural appearance of volume
preserving diffeomorphisms, which can be given a brane
interpretation~\cite{volpres}, is in fact a general feature of the
spacetime symmetries induced by noncommutative gauge theory~\cite{LSred}.

A generic transformation of the form (\ref{deltaVf}) is parametrized by rank
$d-2$ tensors $\chi^{~}_{i_1\cdots i_{d-2}}(x)$ as
\be
V=\epsilon^{ii_1\cdots i_{d-1}}\,\partial_{i_{d-1}}
\chi^{~}_{i_1\cdots  i_{d-2}}(x)~\partial_i \ .
\label{genvolpres}\ee
The subalgebra ${\rm sdiff}(\real^d)$ of symplectic diffeomorphisms comes from
taking the vector fields $V=V_F$, where
\beq
V_F=\theta^{ij}\,\partial_iF~\partial_j
\label{VFdef}\eeq
for $F\in\alg_0$ generates the canonical transformation
\beq
V_F(f)=\{f,F\}^{~}_\theta \ .
\label{sympldiff}\eeq
Here
\be
\{f,g\}^{~}_\theta=\theta^{ij}\,\partial_if\,\partial_jg
\label{Poissonbracket}\ee
is the Poisson bracket on the algebra $\alg_0$, and the operators (\ref{VFdef})
generate the Poisson-Lie algebra of $\real^d$,
\beq
[V_F,V_G]=V_{\{F,G\}^{~}_\theta} \ .
\label{PoissonLie}\eeq
We will see in fact that if we identify noncommutative $\real^d$ with the
quantum mechanical phase space of a point particle, then the transformations
\eqn{VFdef} induce a Lie algebra of quantum deformed canonical transformations.
Similar algebras have been studied within the same context
in~\cite{GozziReuter}.

In two dimensions the canonical transformations coincide with the area
preserving diffeomorphisms of spacetime, but for $d>2$ they form a proper
subalgebra of the algebra of volume preserving diffeomorphisms. Inner
automorphisms of the algebra $\alg_\theta$ can only generate point
transformations which correspond to symplectic diffeomorphisms. Because of the
derivation property (\ref{partialinner}), we necessarily have
$i\,[x_i,\lambda]^{~}_\star=-\theta_i^{~j}\,\partial_j\lambda$, and so any
gauge function $\lambda$ generates a canonical transformation of the spacetime
coordinates. In fact, it is well-known that to first order in $\theta_{ij}$
symplectic diffeomorphisms can be realized as inner automorphisms on a Moyal
space. This can be seen explicitly by examining the first few terms of the
series expansion \eqn{commrs} in $\theta_{ij}$, which for $n^i,m^i>2$ yields
\bea
\Bigl[T_{\vec n}\,,\,T_{\vec m}\Bigr]^{~}_\star&=&
\sum_{i\neq j}i\,\theta_{ij}\,n^i\,m^j~T_{\vec n+\vec m-\vec e_i-\vec e_j}
\nn\\&&+\,\sum_{i\neq j}\frac{(i\,\theta_{ij})^3}6\,
\left[n^j(n^j-1)(n^j-2)m^i(m^i-1)(m^i-2)\right.\nn\\&&-\,
n^i(n^i-1)(n^i-2)m^j(m^j-1)(m^j-2)\nn\\
&&+\left.3\,n^in^jm^im^j\left((n^j-1)(m^i-1)-(n^i-1)(m^j-1)\right)\right]~
 T_{\vec n+\vec m-3(\vec e_i+\vec e_j)}\non&&+\,O\left(\theta_{ij}^5\right) \ .
\label{truncated}\eea
It is easy to see that the truncation of the expansion (\ref{truncated}) to
order $\theta_{ij}$ satisfies the Jacobi identity. The $\star$-commutation
relations to this order close to a Lie algebra which approximates the full
gauge algebra $u(\alg_\theta)$. In fact, in each commuting skew-block $a$ we
can define the operators
\beq
{\cal L}_{\vec n}^{(a)}=T_{\vec n+\underline{1}}^{(a)} \ , ~~ n^i\geq-1 \ ,
\label{calLTdef}\eeq
and find that the first line of (\ref{truncated}) realizes the commutation
relations of the $W_{1+\infty}$ algebra
\beq
\left[{\cal L}_{\vec n}^{(a)}\,,\,{\cal L}_{\vec m}^{(a)}\right]_\star
=2i\,\Bigl((\vec n+\underline{1}\,)\wedge(\vec m+\underline{1}\,)\Bigr)^{(a)}~
{\cal L}_{\vec n+\vec m}^{(a)}+O\left(\theta_a^3\right) \ .
\label{W1infty}\eea
We have defined the antisymmetric bilinear form
\beq
\vec n\wedge\vec m=\frac12\,n^i\,\theta_{ij}\,m^j
\label{wedgedef}\eeq
corresponding to the symplectic structure of $\real^d$. Regarding
noncommutative $\real^d$ as a symplectic space, the first order
$\star$-commutator is in fact the Poisson bracket~\cite{BFFLS}. This is not
true if one truncates the $\star$-commutators at higher orders in
$\theta_{ij}$, because while a generic bracket which is a deformation of the
Poisson bracket (such as the Moyal bracket) generates a Lie algebra, such
deformations are either of order 0 or contain infinitely many terms. Note that
the presence of the Tomita involution $J$ effectively doubles the
$W_{1+\infty}$ symmetry above to two commuting copies~\cite{GozziReuter}, so
that the total symmetry algebra is $W_{1+\infty}\otimes W_{1+\infty}^\vee$ with
$W_{1+\infty}^\vee=J\,W_{1+\infty}\,J^{-1}$.

Thus, in the limit $\theta_{ij}\to 0$ with the rescaled generators
$\theta_i^{~j}\,T_{\vec n}$ finite, the noncommutative gauge algebra
truncates to the Poisson-Lie algebra of $\real^d$ defined by the Poisson
bracket (\ref{Poissonbracket}), i.e. $[T_{\vec n},T_{\vec
m}]^{~}_\star=i\,\{T_{\vec n},T_{\vec m}\}^{~}_\theta$ as $\theta_{ij}\to0$. In
this limit, the operators $T_{\vec n}$ are the generators
\beq
W_{\vec n}\equiv V_{T_{\vec n}}=\sum_{i\neq j}n^i\,\theta_{ij}\,
T_{\vec n-\vec e_i}~\partial_j
\label{Vvecn}\eeq
of symplectic diffeomorphisms which preserve the symplectic geometry of
$\real^d$ (equivalently the Poisson bracket (\ref{Poissonbracket})) and the
canonical transformation is an inner automorphism of the algebra $\alg_0$.
We have found therefore that the gauge algebra, leading to generic volume
preserving transformations (\ref{avgtransf}), is a {\em deformation} of the
Poisson-Lie algebra on a commutative space. In contrast, for $d>2$ the Lie
algebra of volume preserving diffeomorphisms of $\real^d$ cannot be
deformed. The two sorts of gauge transformations agree in the infrared
limit of the noncommutative gauge theory, but differ drastically in the
ultraviolet regime where the effects of noncommutativity become important.
Indeed, because of the UV/IR mixing property of noncommutative field
theories~\cite{UVIR}, the equivalence between the Moyal and Poisson brackets in
the commutative limit will cease to be exact at the quantum level. The
deformation of diffeomorphisms that occurs here is analogous to the
deformations of spacetime symmetries that occur in string theory due to a
finite Regge slope $\alpha'$, and indeed the noncommutativity scale plays a
role analogous to the string scale~\cite{IIKK2}, being both a regulator and a
source of non-locality in the theory. One of the intents of the ensuing
analysis is
to give a more precise geometric meaning to this deformation, and to describe
in exactly what sense it can be understood as a diffeomorphism of flat space
$\real^d$. We remark here that the geometrical meaning of the higher order
terms in the expansion of \eqn{commrs} in $\theta_{ij}$ is unclear. The
$O(\theta_{ij}^3)$ correction of \eqn{truncated} is known to correspond to a
symplectic connection, i.e. a connection of a complex line bundle over
$\real^d$ which preserves its Poisson structure.

\subsection{Algebraic Description}

In this subsection we will interpret the gauge Lie algebra $u(\alg_\theta)$ of
noncommutative electrodynamics as an appropriate completion of the infinite
unitary Lie algebra $u(\infty)$. We have already seen in
section~\ref{relations} that gauge transformations can be lifted to inner
automorphisms on a given $\alg_\theta$-module $\hil$. The unitary operators on
$\hil$ are intimately related to the automorphisms of the non-unital
$C^*$-algebra ${\cal K}(\hil)$ of compact operators acting on the Hilbert
space, i.e. the completion of
the algebra of finite-rank operators on $\hil$ in the operator-norm
\beq
\|F\|^{~}_\infty=\sup_{(\psi,\psi)_\hil\leq1}\,\sqrt{(F\psi,F\psi)_\hil} \ ,
\label{opnorm}\eeq
where in general $F$ is a bounded linear operator on $\hil$ and
$(\cdot,\cdot)_\hil$ denotes the inner product on the Hilbert space. The
physical relevance of subalgebras of compact operators lies in the content of
the
Stone-von~Neumann-Mackey theorem~\cite{Rieffel}, which asserts that $\hil$ is
unitarily equivalent to a direct sum of Schr\"odinger $\alg_\theta$-modules
(\ref{QMhil}). To determine the gauge algebra $u(\alg_\theta)$, we will
thereby focus our attention on the irreducible Schr\"odinger representation
of the algebra $\alg_\theta$. The image under the Weyl map $\Omega$ of the
algebra ${\cal S}(\real^d)$ of Schwartz functions is contained in the algebra
of compact operators on the quantum mechanical Hilbert space $\hil_{\rm Q}$
\footnote{\baselineskip=12pt Generally, from (\ref{Trint}) it follows that if
$f\in L^1(\reals^d,\dd^dx)$, then $\Omega(f)$ is a trace-class operator on
$\hil$ and hence is compact. If $f\in L^2(\reals^d,\dd^dx)$ and $\hil$ is
the Hilbert space (\ref{QMhil}) of quantum mechanics, then $\Omega(f)$ is a
Hilbert-Schmidt operator on $\hil_{\rm Q}$ \cite{AgarwalWolf}. For further
discussion of larger classes of functions which have compact Weyl
transforms, see \cite{Howe}.}. Completing the algebra ${\cal A}_\theta$ in
the $L^\infty$-norm (\ref{supnorm}) yields an algebra whose $\Omega$-image
is the $C^*$-algebra of compact operators ${\cal K}(\hil_{\rm Q})$ (see for
example Lemma~3.41 of \cite{Ticos}). It is important at this point to
notice that while the norm completion of the commutative algebra ${\cal
A}_0$ is the $C^*$-algebra $C(\real^d)$ of continuous functions on
$\real^d\to\complex$, the $C^*$-completion of ${\cal A}_\theta$ is {\em
not} the algebra $C(\real^d)$ equipped with the $\star$-product. In fact,
some smoothness restrictions on the functions are required in order that
their Weyl maps yield bounded operators \cite{completion}. This is
certainly true for the Schwartz space ${\cal S}(\real^d)$ of functions that
we are presently working with. The sorts of generalized functions
comprising the $C^*$-completion of the algebra $\alg_\theta$ are not known
in general.

Nevertheless, for our purposes it will suffice to know that the image of the
completion of $\alg_\theta$ under the Weyl map $\Omega$ is the algebra of
compact operators in the case $\hil=\hil_{\rm Q}$. Furthermore, the Hilbert
space $\hil_{\rm Q}\otimes\hil_{\rm Q}^\vee$, regarded as the space of
square-integrable functions of the complex coordinates $z$ and $z^*$, is
precisely the algebra of Hilbert-Schmidt operators on the Hilbert space
$\hil_{\rm Q}$ \cite{AgarwalWolf}. It represents the joining of open string
endpoints, and it is dense (in the operator-norm topology) in ${\cal
K}(\hil_{\rm Q})$. Generally, given $U\in U(\hil)$, there is the natural
automorphism $\imath_U$ of ${\cal K}(\hil)$. The map $U\mapsto\imath_U$ defines
a continuous homomorphism (in the operator-norm topology) of $U(\hil)$ onto
${\rm Aut}({\cal K}(\hil))$ with kernel $U(1)$ corresponding to the phase
multiples of the identity operator $\id_\hil$. This identifies the automorphism
group
\beq
{\rm Aut}\Bigl({\cal K}(\hil)\Bigr)=PU(\hil) \ ,
\label{Autcompactgen}\eeq
where $PU(\hil)=U(\hil)/U(1)$ is the group of projective unitary
automorphisms of the Hilbert space $\hil$. Note that the defining
representation of ${\cal K}(\hil)$ on $\hil$ is, up to unitary equivalence,
the only irreducible representation of the $C^*$-algebra of compact
operators.

For most of the analysis which follows in this subsection it will suffice to
consider the Fock space (\ref{Fock}) of a given skew-block $a$ of the
antisymmetric tensor $\theta_{ij}$. At the end of the analysis we can then
easily stitch the skew-blocks together to produce the final result. We will
thereby effectively work in $d=2$ dimensions and drop the $a$ indices on all
quantities for the most part. We shall therefore start by analysing the algebra
${\cal K}(\ell^2(\zed_+))$ of compact operators on the Fock space
$\ell^2(\zed_+)=\bigoplus_{n\geq0}\complex|n\rangle$. It is the operator-norm
closure
\beq
{\cal K}\Bigl(\ell^2(\zed_+)\Bigr)={\cal C}_\infty^*
\Bigl(\mat_\infty(\complex)\Bigr)
\label{calKclosure}\eeq
of the algebra $\mat_\infty(\complex)$ of finite-rank operators acting on
$\ell^2(\zed_+)$. The latter algebra is the inductive $N\to\infty$ limit
\beq
\mat_\infty(\complex)=\bigcup_{N=1}^\infty\mat_N(\complex)
\label{matinfty}\eeq
of finite-dimensional, $N\times N$ matrix algebras $\mat_N(\complex)={\cal
K}(\complex^N)$. As a set, (\ref{matinfty}) is made of coherent sequences with
respect to the natural system of embeddings
\bea
\mat_N(\complex)&\hookrightarrow&\mat_{N+1}(\complex)\non
M&\longmapsto&\pmatrix{M&0\cr0&0\cr} \ .
\label{matNemb}\eea
The unitary group of the algebra $\mat_N(\complex)$ is just the usual $N\times
N$ unitary group $U(N)$, and the map $\imath:U(N)\to{\rm
Inn}(\mat_N(\complex))$ has kernel $\ker\imath=U(1)$. The group of
finite-dimensional inner automorphisms is thereby given as
\beq
{\rm Inn}\Bigl(\mat_N(\complex)\Bigr)=U(N)/U(1)=SU(N)/\zed_N \ .
\label{InnmatN}\eeq
The large $N$ completion of (\ref{InnmatN}) then coincides with the
automorphism group (\ref{Autcompactgen}) for the Hilbert space
$\hil=\ell^2(\zed_+)$. In the following we will not deal with any global
aspects of the gauge groups and focus only on the infinitesimal gauge
transformations. From the preceding arguments it is then clear that the
gauge algebra of (\ref{calKclosure}) is the operator-norm closure
\beq
u\left[{\cal K}\Bigl(\ell^2(\zed_+)\Bigr)\right]={\cal C}_\infty^*
\Bigl(u(\infty)\Bigr) \ ,
\label{calKunitary}\eeq
where
\beq
u(\infty)=\bigcup_{N=1}^\infty u(N)
\label{uinfty}\eeq
is the infinite-dimensional Lie algebra of finite-rank Hermitian operators
on Fock space.

The result (\ref{calKunitary}) illustrates how the $U(1)$ noncommutative
gauge theory contains all non-abelian unitary gauge groups. In the large
$N$ limit of (\ref{InnmatN}) leading to (\ref{Autcompactgen}), we replace
the global center subgroup $\zed_N$ of $SU(N)$ by the phase group $U(1)$.
The {\it local} gauge group, or equivalently the gauge algebra, is thereby
intimately related to the infinite-dimensional Lie algebra $su(\infty)$ of
traceless finite-rank Hermitian operators. In fact, the description of this
subsection can be related to the standard Cartan basis of
$su(\infty)$~\cite{Kac} by using the presentation of the gauge algebra of
section 3.2. For this, we
introduce the step operators
\beq
\hat E_{\vec n}=|n^1\rangle\langle n^2|=
\frac12\,\left(\hat\Sigma_{\vec n}^-+i\,\hat\Sigma_{\vec n}^+\right)
\label{hatEdef}\eeq
for each two-vector $\vec n=(n^1,n^2)$ of non-negative integers. The operators
(\ref{hatEdef}) are orthonormal in the standard inner product on the space
$\mat_\infty(\complex)$,
\beq
\Tr_\infty\left(\hat E_{\vec n}^\dagger\,\hat E_{\vec m}\right)
=\delta_{\vec n\,\vec m} \ ,
\label{hatEortho}\eeq
and they thereby span the linear space of finite-rank operators on
$\ell^2(\zed_+)$. The Cartan subalgebra of $su(\infty)$ is then taken to be
generated by the traceless diagonal operators
\beq
\hat H_n=\hat E_{\underline{n}}-\hat E_{\underline{n+1}} \ , ~~ n\in\zed_+ \ .
\label{hatHdef}\eeq
The simple root vectors $\vec\alpha_n$ have components defined by
\beq
\left[\hat H_m\,,\,\hat E_{\underline{n}+\vec e_2}\right]=\left(
\vec\alpha_n\right)_m\,\hat E_{\underline{n}+\vec e_2} \ ,
\label{simpledef}\eeq
so that
\beq
\left(\vec\alpha_n\right)_m=-\delta_{n-1,m}+2\,\delta_{nm}-\delta_{n+1,m} \ .
\label{simplecomps}\eeq
The Fock space representation of the infinite-dimensional Lie algebra
$su(\infty)$ may thereby be given as the semi-infinite Dynkin diagram
\unitlength=1.00mm
\linethickness{0.4pt}
\begin{equation}
\begin{picture}(60.00,5.00)
\thinlines
\put(2.00,2.00){\circle{4.00}}
\put(4.00,2.00){\line(1,0){6.00}}
\put(12.00,2.00){\circle{4.00}}
\put(14.00,2.00){\line(1,0){6.00}}
\put(22.00,2.00){\circle{4.00}}
\put(24.00,2.00){\line(1,0){6.00}}
\put(32.00,2.00){\circle{4.00}}
\put(34.00,2.00){\line(1,0){6.00}}
\put(42.00,2.00){\circle{4.00}}
\put(44.00,2.00){\line(1,0){4.00}}
\put(50.00,2.00){$\dots \ . $}
\end{picture}
\label{Dynkin}\end{equation}
The Lie algebra $u(\alg_\theta)$ is nothing but a ``Schwartzian'' version of
$su(\infty)$, with which it shares the same basis.

We must remember however that we are dealing here with infinite dimensional
algebras, and that therefore the identifications have to be made with great
care. While generally for $su(\infty)$ what is meant is the space of
arbitrarily large but {\em finite} matrices, we are using Schwartzian but
otherwise infinite combinations of the basis elements \eqn{hatEdef}.
If one restricts the Lie algebras to {\em finite} combinations of the
generators, then the two presentations that we have constructed in section~3
would not be representations of the same algebra because the change of basis
described in
section~3.3 involves infinite (Schwartzian) series. Moreover, there are
infinitely many (non-mutually pairwise isomorphic) versions of $su(\infty)$
which depend on the way that the large $N$ limit is taken~\cite{HoppeSchaller}.
Indeed, there are many distinct algebras that can be obtained from
$\mat_N(\complex)$ by taking more complicated embeddings than the simplest,
canonical one (\ref{matNemb}) (see~\cite{largeNnc} for the example of the
noncommutative torus). The connections between $su(\infty)$ and the area
preserving diffeomorphisms of a two-dimensional torus are of a similar kind. In
that case the two Lie algebras are not isomorphic \cite{BHSS}. However, their
difference lies in the different limits, or equivalently in the high frequency
components, and for a variety of physical applications the difference will not
be relevant from the point of view of infinitesimal gauge transformations. This
discussion clarifies the explicit connection between noncommutative $U(1)$
gauge theories and $su(\infty)$, and hence matrix models.

\subsection{Geometrical Description}

We will now pass from an algebraic description of the gauge algebra
$u(\alg_\theta)$ to a more geometrical one. For this, we note that the Weyl
operators (\ref{Weyl}) should really be regarded as elements of the Heisenberg
group algebra, which is locally generated by the unitary translation operators
\beq
\hat{\cal T}(\xi)=\exp\Bigl(i\,\xi\cdot\hat x\Bigr)=\sum_{\vec n\in\zeds_+^d}
\frac{(-1)^{|\vec n|/2}\,(\xi^1)^{n^1}\cdots(\xi^d)^{n^d}}
{n^1!\cdots n^d!}~\hat T_{\vec n} \ ,
\label{calTxi}\eeq
where $\xi^i\in\real$. They enjoy the properties
\bea
\hat{\cal T}(\xi)^\dagger&=&\hat{\cal T}(-\xi) \ , \label{calTadj}\\
\hat{\cal T}(\xi)\,F(\hat x)\,\hat{\cal T}(\xi)^\dagger&=&F(\hat x+\theta\,
\xi) \ , \label{hatTtransl}\\\hat{\cal T}(\xi)\,\hat{\cal T}(\xi')&=&
\e^{-i\,\xi\wedge\xi'}\,\hat{\cal T}(\xi+\xi') \ ,
\label{translgroup}\eea
where $F\in\alg_\theta(\hil)$. In the skew-diagonalization we may use
(\ref{hatxhatzrel}) to equivalently write
\beq
\hat{\cal T}(\xi)=\exp\left(i\,
\xi_z^a\,\hat z_a+i\,\xi_z^{a\,*}\,\hat z_a^\dagger\right)
\label{calTzrep}\eeq
as operators on the quantum mechanical Hilbert space $\hil_{\rm Q}$, where
\beq
\xi_z^a=\sqrt{\frac{\theta_a}2}\,\Bigl(\xi^{2a-1}-i\,\xi^{2a}\Bigr) \ .
\label{xiza}\eeq
We may thereby associate to the translation operators the coherent states
\beq
|\xi_z\rangle=\hat{\cal T}(\xi)|\vec0\,\rangle \ ,
\label{coherent}\eeq
where $|\vec0\,\rangle=|0\rangle\otimes\cdots\otimes|0\rangle$ is the $\frac
d2$-fold Fock vacuum state of the Hilbert space (\ref{QMhil}). We recall from
section~2.2 that the subspace spanned by the coherent states (\ref{coherent})
is precisely one of the irreducible components of the $\alg_\theta$-module
$\hil_{\rm M}$ of fundamental matter fields.

{}From (\ref{translgroup}) we may infer the commutation relations
\beq
\left[\hat{\cal T}(\xi)\,,\,\hat{\cal T}(\xi')\right]=-2i\,\sin\left(
\xi\wedge\xi'\right)~\hat{\cal T}(\xi+\xi') \ .
\label{calTcomm}\eeq
The advantage of passing to the Heisenberg-Weyl group is that it admits
(in the compact case) finite-dimensional representations which can be used
to map the unitary group representations above to geometric ones. Namely,
when $\theta=\frac MN$, with $M,N\in\zed_+$ coprime, and the
Fourier momenta $\xi$ are restricted to lie on the lattice $\zed_+^d$, then
there is an $N$-dimensional unitary representation of the commutation
relations (\ref{calTcomm}). This restriction on $\theta$, once a large $N$
limit is taken, does not affect expectation values of operators nor any
physically measurable quantities \cite{largeNnc}. The restriction on the
momenta means that the space we are effectively considering is a
(noncommutative) torus. Since we are concerned here only with infinitesimal
gauge transformations, one might think that the change from the compact
to the non-compact case will not alter the structure of the Lie algebra
$u(\alg_\theta)$. This is not precisely correct, because in noncommutative
quantum field theory there is an interplay between local (ultraviolet) and
global (infrared) characteristics. We shall see in fact that the gauge Lie
algebra we find in this case is not exactly the one that we found in the
previous subsection.

For simplicity we shall assume that $N$ is odd. For each $\vec
n=(n^1,n^2)\in\zed_+^2$, we introduce the $N\times N$ unitary unimodular
matrices
\beq
\hat{\cal W}^{(N)}_{\vec n}=\omega^{n^1n^2/2}~\left(\hat\Gamma_{\rm c}
\right)^{n^1}\,\left(\hat\Gamma_{\rm s}\right)^{n^2} \ ,
\label{hatJNdef}\eeq
where $\omega=\e^{4\pi i\,M/N}$ is an $N^{\rm th}$ root of unity. The matrices
$\hat\Gamma_{\rm c}$ and $\hat\Gamma_{\rm s}$ generate the Weyl-'t~Hooft
algebra
\beq
\hat\Gamma_{\rm c}\,\hat\Gamma_{\rm s}=\omega\,\hat\Gamma_{\rm s}
\,\hat\Gamma_{\rm c} \ ,
\label{WeyltHooft}\eeq
which, up to a gauge transformation $\hat\Gamma_i\mapsto
U\,\hat\Gamma_i\,U^{-1}$ with $U\in SU(N)$, may be uniquely represented by the
usual clock and cyclic shift operators for $SU(N)$,
\bea
\hat\Gamma_{\rm c}&=&\sum_{n=0}^{N-1}\omega^n\,|n\rangle
\langle n|~=~-\sum_{n\geq0}\omega^n\,\sum_{m=0}^{n-1}\hat H_m \ , \non
\hat\Gamma_{\rm s}&=&\sum_{n=0}^{N-1}\,|n+1\,({\rm mod}\,N)
\rangle\langle n|~=~
\sum_{n\geq0}\hat E_{\underline{n}+\vec e_1\,({\rm mod}\,N)} \ .
\label{Gammarep}\eea
The finite-rank operators (\ref{hatJNdef}) satisfy the unitarity condition
\beq
\hat{\cal W}^{(N)\,\dagger}_{\vec n}=\hat{\cal W}^{(N)}_{-\vec n} \ ,
\label{hatJunitary}\eeq
and, because $(\hat\Gamma_i)^N=\id_N$, they are periodic in $n^i$ modulo $N$,
\beq
\hat{\cal W}^{(N)}_{\vec n+N\vec e_i}=\hat{\cal W}^{(N)}_{\vec n} \ , ~~ i=1,2
\ .
\label{hatJperiod}\eeq
The property (\ref{hatJperiod}) implies that there are only $N^2$ distinct
operators (\ref{hatJNdef}) corresponding to the values $\vec
n\in\zed_N^2$. Furthermore, they are orthonormal in the standard inner
product on the space of complex matrices,
\beq
\Tr_N\left(\hat{\cal W}_{\vec n}^{(N)\,\dagger}\,\hat{\cal W}^{(N)}_{\vec m}
\right)=\delta_{\vec n\,\vec m} \ .
\label{hatJortho}\eeq
It follows that the underlying linear space of the $C^*$-algebra
$\mat_N(\complex)$ is spanned by the matrices (\ref{hatJNdef}). This is known
as the Weyl basis for the Lie algebra $gl(N,\complex)$.

In fact, from (\ref{hatJortho}) it follows that the $N^2-1$ matrices $\hat
{\cal W}_{\vec n}^{(N)}$, $\vec n\neq\vec0$, form a complete set of traceless
unitary matrices. By taking real and imaginary combinations, they
therefore span the Lie algebra $su(N)$. From (\ref{WeyltHooft}) we also
find that these operators close projectively under multiplication,
\beq
\hat{\cal W}_{\vec n}^{(N)}\,\hat{\cal W}_{\vec m}^{(N)}=
\omega^{\vec m\wedge\vec n}\,
\hat{\cal W}_{\vec n+\vec m\,({\rm mod}\,\underline{N}\,)}^{(N)} \ ,
\label{hatJmult}\eeq
which specifies an additive grading structure. The relation (\ref{hatJmult})
further leads to an explicit form for the $su(N)$ structure constants in the
Weyl basis through the commutation relations of the Fairlie-Fletcher-Zachos
trigonometric algebra~\cite{FairlieZachos}
\beq
\left[\hat{\cal W}_{\vec n}^{(N)}\,,\,\hat{\cal W}_{\vec m}^{(N)}\right]=\frac
{i\,N}{2\pi M}\,
\sin\left(\frac{4\pi M}N\,\vec n\wedge\vec m\right)\,
\hat{\cal W}_{\vec n+\vec m\,({\rm mod}\,\underline{N}\,)}^{(N)} \ ,
\label{hatJcomms}\eeq
where we have rescaled the operators $\hat{\cal W}^{(N)}_{\vec
n}\mapsto\frac N{4\pi M}\,\hat{\cal W}^{(N)}_{\vec n}$. The explicit
transformation between the Weyl and Cartan bases of $su(N)$ can be obtained
from
(\ref{hatEortho}), (\ref{Gammarep}) and (\ref{hatJortho}) to give
\bea
\hat{\cal W}_{\vec n}^{(N)}&=&-\frac N{4\pi M}\,\omega^{n^2(n^1+n^2+1)/2}
\,\sum_{m=0}^{N-1}\omega^{m(n^1+n^2)}\,\hat E_{\underline{m}-n^2\vec e_2\,
({\rm mod}\,N)} \ , \non\hat E_{\vec n\,({\rm mod}\,\underline{N}
\,)}&=&-\frac{4\pi M}N
\,\omega^{(n^1-n^2)(n^1-2n^2-1)/2}\,\sum_{m=0}^{N-1}\omega^{m(n^1-2n^2)/2}
\,\hat{\cal W}_{m\vec e_1+(n^1-n^2)\vec e_2}^{(N)} \ .
\label{hatEJchange}\eea

We see therefore that the finite-dimensional operators (\ref{hatJNdef})
formally possess the same algebraic properties as the translation generators
(\ref{calTxi}), and can thereby be thought of as a certain $N\times N$
approximation to them. Passing to the inductive large-$N$ limit yields
generators $\hat{\cal W}_{\vec n}^{(\infty)}$, $\vec n\neq\vec0$, of
$su(\infty)$ which from (\ref{hatJcomms}) satisfy the commutation relations for
the classical $W_\infty$ algebra
\beq
\left[\hat{\cal W}_{\vec n}^{(\infty)}\,,\,\hat{\cal W}_{\vec m}^{(\infty)}
\right]=2i\,\vec n\wedge\vec m~\hat{\cal W}_{\vec n+\vec m}^{(\infty)} \ .
\label{hatJinftycomms}\eeq
Again there are actually two commuting copies of this $W_\infty$ algebra, so
that the total symmetry algebra is $W_\infty\otimes W_\infty^\vee$. The
$W_\infty$ component acts separately on each Landau level of the oscillator
space (\ref{Fock}), while $W_\infty^\vee$ mixes the different levels but acts
in a simple way on the coherent states (\ref{coherent}).

The algebra (\ref{hatJinftycomms}) can be identified with the Lie algebra of
the vector fields
\beq
\hat{\cal W}_{\vec n}^{(\infty)}=i\,V_{{\cal T}_{\vec n}}={\cal T}_{\vec n}~
\vec n\wedge\partial \ ,
\label{VcalT}\eeq
where ${\cal T}_{\vec n}(x)=\e^{i\,\vec n\cdot x}$ are the complete set of
harmonics on a two-dimensional square torus. However, from the point of
view of infinitesimal diffeomorphisms, this torus basis is simply a matter
of convenience. Locally, the canonical transformations generated by
(\ref{VcalT}) are simply those which preserve the symplectic two-form $\dd
x\wedge\dd x$, and this property holds whether we are speaking of the torus
or all of $\real^d$. This is evident in particular from the natural local
isomorphism
between the $W_{1+\infty}$ algebra (\ref{W1infty}) and the $W_\infty$ algebra
(\ref{hatJinftycomms}). The global group structure is a somewhat more subtle
issue which will not be dealt with in this paper. From (\ref{hatEJchange}) we
find the explicit relationship between the symplectomorphism generators
(\ref{VcalT}) in the torus basis and the Cartan basis of $su(\infty)$ given
in the previous subsection. Again, it should be stressed that the large $N$
limit of (\ref{hatJcomms}) leading to (\ref{hatJinftycomms}) implies ignoring
periodicity and the boundaries of the Brillouin zone in momentum space. This
has no bearing however on the local aspects of the problem.

By including the local translation generators $\hat{\cal W}_{\vec0}^{(\infty)}$
which are excluded from (\ref{VcalT}) (the torus has only global translational
isometries), we find an extra $U(1)$ symmetry on the plane and hence the Lie
algebra $u(\infty)$. At the infinitesimal level this extra $u(1)$ factor, which
emerges from the infrared properties of the torus, is somewhat ``sterile''
because it corresponds to a mode which $\star$-commutes with all fields. It can
however have consequences for the global properties of the theory. We thereby
find an explicit relationship between the infinite unitary Lie algebra
$u(\infty)$ and the symplectomorphism algebra, and hence with the gauge algebra
(\ref{calKunitary}). Let us emphasize again that the explicit connection made
in this section between the noncommutative gauge algebra and the Lie algebra
(\ref{uinfty}) holds strictly only in the truncation to finite rank operators.
In particular, because of the boundary effects described above, there is no
immediate isomorphism with the group of symplectomorphisms of $\real^d$.
Nevertheless, all of these algebras complete in norm to the same subalgebra of
compact operators if the maps are chosen appropriately.

These properties evidently all persist when gluing the skew-blocks together
again to produce the full Hilbert space (\ref{QMhil}), giving the unitary gauge
algebra of the entire noncommutative space, $u(\alg_\theta)={\cal
C}_\infty^*(u(\infty)\otimes\real^{d/2})\cong{\cal C}_\infty^*(u(\infty))$. The
main conclusion of this section is then that the gauge algebra of Yang-Mills
theory on a Moyal space is a deformation of the Poisson-Lie algebra of
$\real^d$ which, for the dense class of functions we are considering, is
locally isomorphic to $u(\infty)$ in the norm closure,
\beq
u(\alg_\theta)={\cal C}_\theta^*\Bigl({\rm
sdiff}(\real^d)\Bigr)\cong{\cal C}_\infty^*\Bigl(u(\infty)\Bigr) \ .
\label{gaugesdiff}\eeq
Here ${\cal C}_\theta^*$ denotes the $C^*$-completion for the Moyal
bracket, which defines a subspace of generalized functions on $\real^d$
whose Moyal brackets induce the appropriate deformation of the
symplectomorphism algebra ${\rm sdiff}(\real^d)$. The identification
(\ref{gaugesdiff}) is very natural in light of the intimate relationship
that exists between large $N$ matrix models and noncommutative gauge theory
via the Eguchi-Kawai reduction~\cite{AMNS1,AMNS2,AIIKKT}. Indeed, reduced
models transmute spacetime degrees of freedom into internal matrix ones. It is
for this reason, for example, that M2-brane dynamics arise in Matrix
theory~\cite{BFSS}, whereby the symplectic symmetry appears as a discretization
of the residual gauge symmetry of the 11-dimensional supermembrane~\cite{dWHN}.
This symmetry is also related to the equivalence between large $N$ reduced
models and the Schild model~\cite{IKKT}. The generic connection with volume
preserving diffeomorphisms is at the very heart of the way that noncommutative
gauge theory effectively encodes the target space symmetries of
D-branes~\cite{LSred}. The identification (\ref{gaugesdiff}) has also been
noted directly in the context of D-brane field theory in~\cite{CNCrel,HKL}.

\newsection{Spacetime Symmetry Algebras}

One of the most interesting features of gauge theory on a noncommutative space
is the interplay between gauge and spacetime transformations. It is therefore
interesting to study subalgebras of the gauge algebra $u(\alg_\theta)$ and to
see what sorts of geometric interpretations they admit. There can in fact be
situations in which the noncommutative gauge symmetry is broken down to a
subalgebra of $u(\alg_\theta)$, due to an action functional which has a minimum
that is invariant only under the subalgebra (or due to some other symmetry
breaking mechanism). In this case only the gauge transformations of this
subalgebra are available, and in effect reduce the spacetime. We shall return
to this point in the next section.

In the previous section we saw that local gauge transformations are intimately
related to automorphisms of the algebra ${\cal K}(\hil)$ of compact operators
(with respect to an appropriate Weyl representation). From this feature we
identified $u(\alg_\theta)$ as the closure of the infinite unitary Lie algebra
$u(\infty)$ which contains, in particular, as subalgebras all finite
dimensional unitary gauge algebras $u(N)$. It contains them in fact several
times in various different guises. There are also numerous infinite dimensional
subalgebras of $u(\alg_\theta)$. In this section we shall present a rather
elementary description of some of these subalgebras from the point of view of
their actions on spacetime scalar fields. In the next section we will describe
their interpretations as transformations of the gauge fields. Often we will
describe only the two-dimensional case corresponding to a given skew-block of
the deformation matrix $\theta_{ij}$, the stitching together of the blocks in
the end being straightforward.

The basic idea behind the emergence of Lie subalgebras of $u(\alg_\theta)$ is
as follows. Let $\cal G$ be an abstract $n$-dimensional Lie algebra with
structure constants $f_{ab}^{~~c}$ in a chosen basis. Let ${\cal G}^*$ be its
dual, with corresponding generators $X_1,\dots,X_n$ and Lie bracket
$[\,\cdot\,,\,\cdot\,]_{{\cal G}^*}$,
\beq
[X_a,X_b]_{{\cal G}^*}=f_{ab}^{~~c}\,X_c \ .
\label{calGcomm}\eeq
The Lie algebra $\cal G$ may be realized not only in terms of operators, but
also in terms of functions on Poisson manifolds with the Lie bracket replaced
by the Poisson bracket~\cite{Lichner}. We seek a realization of $\cal G$ as an
$n$-dimensional Lie subalgebra ${\cal G}_\theta$ of the noncommutative gauge
algebra $u(\alg_\theta)$. For this, we will construct a map
$\rho:\real^d\to{\cal G}^*$ whose pullbacks generate the Moyal brackets
\beq
\Bigl[\rho^*(X_a)\,,\,\rho^*(X_b)\Bigr]^{~}_\star=i\,f_{ab}^{~~c}\,\rho^*(X_c)
+O(\theta_{ij}) \ .
\label{rhoMoyal}\eeq
The map $\rho$ need not be a projection, and indeed in many of the cases that
we shall consider we can take $d<\dim{\cal G}^*$. The $O(\theta_{ij})$ terms in
(\ref{rhoMoyal}) will generically arise because, as discussed extensively in
the previous section, the Moyal bracket yields a deformation of the Poisson
bracket of $\real^d$. We will see that the ensuing gauge transformations (inner
automorphisms) yield the anticipated diffeomorphisms of spacetime in these
cases. However, because the gauge algebra (\ref{gaugesdiff}) actually contains
the Lie algebra ${\rm sdiff}(\real^d)$ of canonical transformations, we would
expect to be able to find maps $\rho$ which lead to undeformed representations
of the Lie algebra $\cal G$. We shall find that this is indeed the case. But
because the Moyal bracket modifies non-trivially the $C^*$-completion in
(\ref{gaugesdiff}), the corresponding inner automorphisms will not generate
ordinary diffeomorphisms, nor will they admit a natural geometrical
interpretation in terms of noncommutative gauge transformations. The following
analysis will therefore make the meaning of the deformation encountered in the
previous section somewhat more precise, at least from a geometrical point of
view.

\subsection{Gaussian Algebras}

Let us begin with the simplest illustrative example, which involves the
``momentum'' operators
\beq
P_i=-\sum_{j=1}^d\theta_i^{~j}\,T_{\vec e_j}
\label{Piops}\eeq
that act on the algebra $\alg_\theta$ by the inner
automorphisms\footnote{\baselineskip=12pt As discussed in section~2, elements
of the form (\ref{Piops}) morally only generate inner automorphisms of the ring
$\complexs[[x_1,\dots,x_d]]/{\cal R}_\theta$. In the following we will not be
pedantic about the completions to spaces of Schwartz functions and still refer
to transformations of the form (\ref{Piinner}) as inner automorphisms of the
algebra $\alg_\theta$.}
\beq
i\,[P_i,f]^{~}_\star=\partial_if \ , ~~ \forall f\in\alg_\theta \ ,
\label{Piinner}\eeq
and thereby correspond to spacetime translation operators. They generate the
noncommutative translation algebra
\beq
[P_i,P_j]^{~}_\star=i\,B_{ij} \ ,
\label{NCtranslalg}\eeq
where
\beq
B_{ij}=\delta_{ik}\,\theta_j^{~k}
\label{Bfield}\eeq
is a constant background magnetic field. This central extension of the usual
abelian Lie algebra of translations of $\real^d$ can be understood in terms of
the corresponding group elements (\ref{calTxi}). From (\ref{translgroup}) it
follows that they only generate a projective representation of the abelian
group of translations in spacetime. We can view this as a genuine
representation of a larger group by including the projective phase factors, or
equivalently by augmenting the non-abelian Lie algebra (\ref{NCtranslalg}) by
the central elements appearing on the right-hand side. We recall from
section~3.1 that the $T_{\vec n}$'s constitute a basis for $u(\alg_\theta)$ in
the enveloping algebra of this Lie algebra. As we will see in the next section,
such central extensions have no consequences for the gauge-invariant dynamics
of the field theory.

The translation group can be completed into the Euclidean group $ISO(d)$ in $d$
dimensions. For illustration, we will first demonstrate this for the case of
the Moyal plane $\real^2$ with noncommutativity parameter $\theta$. By defining
the operator
\beq
L_{12}=\frac1{2\theta}\,\Bigl(T_{2\vec e_1}+T_{2\vec e_2}\Bigr) \ ,
\label{L12def}\eeq
we see from (\ref{commrs}) that along with the $\star$-commutation relations
(\ref{NCtranslalg}) we obtain a deformed version of the Euclidean group
$ISO(2)$,
\bea
[L_{12},P_1]^{~}_\star&=&i\,P_2 \ , \nn\\ ~~
[L_{12},P_2]^{~}_\star&=&-i\,P_1 \ .
\label{E2comm}\eea
That $L_{12}$ can be identified as a rotation generator may be seen as follows.
For an arbitrary function
\be
f=\sum_{\vec n\in\zeds_+^2}f_{\vec n}\,T_{\vec n}
\label{farbT}\ee
of the noncommutative algebra $\alg_\theta$, the corresponding inner
automorphism is given from (\ref{commrs}) by
\be
i\,[L_{12},f]^{~}_\star=\sum_{\vec n\in\zeds_+^2}
f_{\vec n}\,\Bigl(n^2\,T_{\vec n
+\vec e_1-\vec e_2}-n^1\,T_{\vec n+\vec e_2-\vec e_1}\Bigr) \ ,
\label{rota}\ee
which is readily seen to be the anticipated form of an infinitesimal rotation
of a scalar field in two dimensions.

The generalization to arbitrary spacetime dimension $d$ is straightforward. We
introduce the $d(d-1)/2$ angular momentum operators
\beq
L_{ij}=\frac12\,\sum_{k=1}^d\Bigl(\theta_i^{~k}\,T_{\vec e_j+\vec e_k}-
\theta_j^{~k}\,T_{\vec e_i+\vec e_k}\Bigr) \ ,
\label{Lijd}\eeq
and find, by using (\ref{commrs}), that along with (\ref{NCtranslalg}) they
satisfy the commutation relations of a deformed, noncommutative $iso(d)$ Lie
algebra,
\bea
[L_{ij},P_k]^{~}_\star&=&\frac i2\,\Bigl(\delta_{ik}\,P_j-\delta_{jk}\,
P_i\Bigr)+\frac i2\,\Bigl(\theta_{ik}\,\theta_j^{~l}-\theta_{jk}\,
\theta_i^{~l}\Bigr)\,P_l \ , \nn\\&&{~~~~}\nn\\
{}~~ [L_{ij},L_{kl}]^{~}_\star&=&\frac i2\,
\Bigl(\delta_{il}\,L_{jk}+\delta_{jk}\,L_{il}-\delta_{ik}\,L_{jl}-\delta_{jl}
\,L_{ik}\Bigr)\nn\\&&+\,\frac i2\,\left[\Bigl(\theta_{jl}\,\theta_k^{~m}-
\theta_{jk}\,\theta_l^{~m}\Bigr)\,L_{im}+\Bigl(\theta_{ik}\,\theta_l^{~m}-
\theta_{il}\,\theta_k^{~m}\Bigr)\,L_{jm}\right.\nn\\&&+\,\Bigl(B_{il}\,
\theta_j^{~m}-B_{jl}\,\theta_i^{~m}\Bigr)\,L_{km}+\Bigl(B_{jk}\,\theta_i^{~m}-
B_{ik}\,\theta_j^{~m}\Bigr)\,L_{lm}\nn\\&&+\left.\Bigl(\delta_{il}\,
\theta_j^{~m}\,\theta_k^{~n}+\delta_{jk}\,\theta_i^{~m}\,\theta_l^{~n}-
\delta_{ik}\,\theta_j^{~m}\,\theta_l^{~n}-\delta_{jl}\,\theta_i^{~m}\,
\theta_k^{~n}\Bigr)\,L_{mn}\right]\nn\\&&+\,\frac i4\,\Biggl[\theta_{ik}\,
T_{\vec e_j+\vec e_k}+\theta_{jl}\,T_{\vec e_i+\vec e_k}-\theta_{il}\,
T_{\vec e_j+\vec e_k}-\theta_{jk}\,T_{\vec e_i+\vec e_l}\Biggr.\nn\\&&+\,
\sum_{m=1}^d\Biggl\{\Bigl(\delta_{il}\,\theta_j^{~m}-\delta_{jl}\,
\theta_i^{~m}\Bigr)\,T_{\vec e_k+\vec e_m}+\Bigl(\delta_{jk}\,\theta_i^{~m}
-\delta_{ik}\,\theta_j^{~m}\Bigr)\,T_{\vec e_l+\vec e_m}\Biggr.\nn\\&&+\,
\Bigl(\delta_{jk}\,\theta_l^{~m}-\delta_{jl}\,\theta_k^{~m}\Bigr)\,
T_{\vec e_i+\vec e_m}+\Bigl(\delta_{il}\,\theta_k^{~m}-\delta_{ik}\,
\theta_l^{~m}\Bigr)\,T_{\vec e_i+\vec e_m}\nn\\&&+\left.\left.\sum_{n=1}^d
\Bigl(B_{ik}\,\theta_j^{~m}\,\theta_l^{~n}+B_{jl}\,\theta_i^{~m}\,
\theta_k^{~n}-B_{il}\,\theta_j^{~m}\,\theta_k^{~n}-B_{jk}\,\theta_i^{~m}\,
\theta_l^{~n}\Bigr)\,T_{\vec e_m+\vec e_n}\right\}\right] \ . \nn\\&&
\label{isodgen}\eea
In both sets of $\star$-commutation relations of (\ref{isodgen}) the first sets
of terms on the right-hand sides represent the standard Lie algebra of the
Euclidean group in $d$ dimensions, while the second terms represent a quantum
deformation of this algebra. However, the $\star$-commutators of angular
momentum operators do not close to a Lie algebra. There is again an extension
of the usual $iso(d)$ algebra, this time by operators which are non-central
elements. We shall see below how this non-central extension may be understood
as a generalization of the projective representation (\ref{translgroup}) to the
full Euclidean group.

The noncommutative translation algebra can also be completed into the
Poincar\'e algebra $iso(1,d-1)$. In the two dimensional case this structure
arises from the generator
\be
K_{12}=\frac1\theta\,T_{\vec e_1+\vec e_2} \ ,
\label{K12def}\ee
which from (\ref{commrs}) yields the $\star$-commutation relations
\bea
[K_{12},P_1]_\star^{~}&=&i\,P_1 \ , \nn\\ ~~
[K_{12},P_2]^{~}_\star&=&-i\,P_2 \ .
\label{Poincare2}\eea
The fact that the gauge algebra of noncommutative Yang-Mills theory contains
deformations of both the Euclidean and Poincar\'{e} algebras is indicative of
the general fact that the signature of a noncommutative space is a delicate
issue from the field theoretical point of view. Indeed, quantum field theories
with noncommuting time coordinate are not unitary and the corresponding
Seiberg-Witten limit in these cases yields a model that only makes sense in
string theory~\cite{NCOS}.

These algebraic results all follow from the fact that these subalgebras are
themselves part of the larger subalgebra of $u(\alg_\theta)$ consisting of
functions which are at most quadratic in the spacetime coordinates. For these
functions, the $\star$-commutator truncates to linear order in the deformation
parameters $\theta_{ij}$, i.e. the Moyal and Poisson brackets coincide, and it
generates a linear canonical transformation of the coordinates $x_i$. Since
\beq
{\rm deg}\,\{f,g\}^{~}_\theta={\rm deg}\,f+{\rm deg}\,g-2
\label{degPoisson}\eeq
for any two polynomial functions $f$ and $g$, it follows that the polynomials
of degree 2 form a Lie subalgebra in the Moyal bracket. Explicitly, from
(\ref{commrs}) we arrive, in addition to (\ref{NCtranslalg}), at the
$\star$-commutation relations
\bea
\left[T_{2\vec e_i}\,,\,T_{2\vec e_j}\right]^{~}_\star&=&4i\,\theta_{ij}\,
T_{\vec e_i+\vec e_j} ~~~~({\rm no}~{\rm sum}~{\rm on}~i,j) \ , \nn\\
\left[T_{2\vec e_i}\,,\,T_{\vec e_j+\vec e_k}\right]^{~}_\star&=&2i\,
\theta_{ij}\,T_{\vec e_i+\vec e_k}+2i\,\theta_{ik}\,T_{\vec e_i+\vec e_j}
{}~~~~({\rm no}~{\rm sum}~{\rm on}~i) \ , \nn\\\left[T_{\vec e_i+\vec e_j}
\,,\,T_{\vec e_k+\vec e_l}\right]^{~}_\star&=&i\,
\theta_{ik}\,T_{\vec e_j+\vec e_l}+i\,\theta_{il}\,T_{\vec e_j+\vec e_k}
+i\,\theta_{jk}\,T_{\vec e_i+\vec e_l}+i\,\theta_{jl}\,
T_{\vec e_i+\vec e_k} \ , \nn\\\left[T_{\vec e_i}\,,\,T_{2\vec e_j}
\right]^{~}_\star&=&2i\,\theta_{ij}\,T_{\vec e_j} ~~~~
({\rm no}~{\rm sum}~{\rm on}~j) \ , \nn\\\left[T_{\vec e_i}\,,\,T_{\vec e_j
+\vec e_k}\right]^{~}_\star&=&i\,\theta_{ij}\,T_{\vec e_k}+i\,\theta_{ik}
\,T_{\vec e_j} \ .
\label{Tquadcomms}\eea
In addition to the deformed Euclidean and Poincar\'e algebras above, there are
a multitude of other subalgebras which can be readily deduced from
(\ref{Tquadcomms}). For instance, in two spacetime dimensions, it is easily
seen from (\ref{Tquadcomms}) that the full set of quadratic operators
\bea
J_3&=&\frac1\theta\,T_{\vec e_1+\vec e_2} \ , \nn\\J_+&=&\frac1{2\theta}
\,T_{2\vec e_2} \ , \nn\\J_-&=&-\frac1{2\theta}\,T_{2\vec e_1}
\label{sl2gens}\eeq
in this case yield a realization of the (undeformed) $sl(2,\real)$ Lie algebra,
\bea
[J_3,J_\pm]^{~}_\star&=&\pm\,2i\,J_\pm \ , \nn\\ ~~ [J_+,J_-]_\star&=&i\,
J_3 \ .
\label{sl2comm}\eeq

The closure of the inner automorphisms at quadratic order can also be seen at
the level of the corresponding group elements
\beq
{\cal Q}(\Delta,\xi)(x)=\det\left(\id_d-\Delta\,\theta\,\Delta\,\theta
\right)^{1/4}\,\exp\left(i\,\Delta^{ij}\,x_i\,x_j+i\,\xi\cdot x\right) \ ,
\label{calQdef}\eeq
where $\xi^i\in\real$ and $\Delta$ is an invertible, symmetric real-valued
$d\times d$ matrix. The Gaussian functions (\ref{calQdef}) are $\star$-unitary,
and we have used the fact that the $\star$-exponential of a quadratic form on
$\real^d$ can be written as an ordinary Gaussian function~\cite{CUZ}. To
compute the generalization of the projective translation representation
(\ref{translgroup}), we use the Fourier integral kernel representation of the
$\star$-product (\ref{defstar})~\cite{Baker}
\beq
(f\star g)(x)=\frac1{\pi^d\,\Bigl|\det(\theta)\Bigr|}\,\int\!\!\!\int\dd^dy~
\dd^dy'~f(y)\,g(y')~\e^{-2i\,(\theta^{-1})^{ij}(x-y)_i(x-y')_j} \ ,
\label{starFourier}\eeq
which follows directly from the Weyl representation
(\ref{Weyl})--(\ref{WWcorr}). By Gaussian integration it yields the group
composition law
\beq
{\cal Q}(\Delta,\xi)\star{\cal Q}(\Delta',\xi')=\Xi\left
(\Delta,\Delta';\xi,\xi'\right)~
{\cal Q}\Bigl(\Lambda(\Delta,\Delta')\,,\,\lambda(\Delta,\Delta';\xi,\xi')
\Bigr) \ ,
\label{quadgrouplaw}\eeq
where
\bea
\Xi(\Delta,\Delta';\xi,\xi')&=&\exp\left[\frac14\,\tr\ln\left(\frac{\id_d
-\Delta'\,\theta\,\Delta'\,\theta}{\id_d+\Delta\,\theta\,
\Delta'\,\theta}\right)+\frac i4\,
\left(\frac1{\Delta}\right)_{ij}~\xi^i\,\xi^j\right.
\non&&-\left.\frac i4\,\left(\frac1{\left(\id_d+\Delta\,\theta\,\Delta'\,\theta
\right)\Delta}\right)_{kl}\,\Bigl(\id_d-\Delta\,\theta\Bigr)_i^{~k}\,
\Bigl(\id_d+\Delta\,\theta\Bigr)_j^{~l}~\xi'^i\,\xi'^j\right] \ , \non
\Lambda(\Delta,\Delta')&=&
-\frac1{\theta\,\Delta\,\theta}+\Bigl(\id_d-\Delta\,\theta\Bigr)\,
\frac1{\left(\id_d+\Delta'\,\theta\,\Delta\,\theta\right)\theta\,\Delta\,
\theta}\,\Bigl(\id_d+\Delta\,\theta\Bigr) \ ,
\non\lambda(\Delta,\Delta';\xi,\xi')&=&-\frac1{\Delta\,\theta}~\xi-
\Bigl(\id_d-\Delta\,\theta\Bigr)\,
\frac1{\left(\id_d-\Delta'\,\theta\,\Delta\,\theta\right)\Delta\,\theta}\,
\Bigl(\id_d+\Delta\,\theta\Bigr)~\xi' \ .
\label{Lambdaeff}\eea

However, at higher orders in the $x_i$'s, whereby the Moyal and Poisson
brackets no longer coincide, this property does not generally hold anymore.
This again owes to the complicated nature of the deformed completion ${\cal
C}_\theta^*$ which defines the gauge algebra (\ref{gaugesdiff}). The
geometrical, diffeomorphism Lie subalgebras arise only in the symplectic limit
whereby the gauge symmetries are associated with isometries of the flat space
$\real^d$. Indeed, the $\star$-product (\ref{defstar}) is only generically
invariant under canonical transformations of the spacetime, since these are
precisely the diffeomorphisms which preserve the Poisson bi-vector
$\partial_\xi\wedge\partial_\eta$ on $\real^d$ that defines (\ref{defstar}).
This is the {\em only} sense in which diffeomorphisms are realizable as inner
automorphisms on a Moyal space.

\subsection{Higher Order Algebras and UV/IR Mixing}

Let us now look at subalgebras generated by more complicated combinations of
the basis elements of $u(\alg_\theta)$. For simplicity we shall again only
consider the two-dimensional case. Now, it is possible to use a more involved
structure and find {\em undeformed} versions of the Lie algebras encountered in
the previous subsection. For example, it is readily checked that the operators
\bea
L_{12}&=&\frac1\theta\,T_{\vec e_2} \ , \nn\\P_1&=&\sum_{n=0}^\infty
\frac{(-1)^n}{(2n)!}\,T_{2n\vec e_1} \ , \nn\\P_2&=&\sum_{n=0}^\infty
\frac{(-1)^n}{(2n+1)!}\,T_{(2n+1)\vec e_1}
\label{E2undef}\eea
generate the standard Lie algebra of the Euclidean group $ISO(2)$, i.e.
$[P_1,P_2]^{~}_\star=0$. The construction can be straightforwardly generalized
to $d$ dimensions. This algebra has been described in \cite{MMVZ} as a
subalgebra of the Poisson-Lie algebra of phase space. In other words, for the
functions (\ref{E2undef}), the Moyal and Poisson brackets again coincide,
showing once more how the geometrical symmetries of noncommutative gauge theory
are tied to the symplectic diffeomorphisms of spacetime.

To understand geometrically what the undeformed operators (\ref{E2undef})
represent, let us compute the corresponding inner automorphisms of the algebra
$\alg_\theta$. Using the translational property (\ref{hatTtransl}), we find
their actions on a function $f\in\alg_\theta$ to be given by
\bea
i\,[L_{12},f]^{~}_\star &=&\del_1 f \ , \nn\\
i\,[P_1,f]^{~}_\star &=&\left(\overrightarrow{\sf d}_1\,f\right)
\star{\cal T}(-\vec e_1)-\left(f\,\overleftarrow{\sf d}_1\right)
\star{\cal T}(\vec e_1) \ ,
\nn\\i\,[P_2,f]^{~}_\star &=&-i\,\left(\overrightarrow{\sf d}_1\,f\right)
\star{\cal T}(-\vec e_1)-i\,\left(f\,\overleftarrow{\sf d}_1\right)
\star{\cal T}(\vec e_1) \ ,
\label{undefinner}\eea
where generally the forward and backward shift operators $\overrightarrow{\sf
d}_i$ and $\overleftarrow{\sf d}_i$ are defined by
\bea
\overrightarrow{\sf d}_i\,f(x)&=&\frac12\,\Bigl(f(x+\theta\,\hat\imath)
-f(x)\Bigr) \ , \nn\\ f(x)\,\overleftarrow{\sf d}_i&=&\frac12\,\Bigl(
f(x)-f(x-\theta\,\hat\imath)\Bigr) \ ,
\label{Deltatheta}\eea
with $\hat\imath$ a unit vector in the $i^{\,\rm th}$ direction of
spacetime. {}From (\ref{undefinner}) we see that this undeformed
representation of the Euclidean algebra acts in an unusual geometric way.
The ``rotation'' $L_{12}$ acts by translations in the $x_1$ direction. In
other words, the representation (\ref{E2undef}) selects a choice of axes in
which it generates non-compact rotations. The two ``translations'' $P_1$
and $P_2$ affect a {\em lattice} displacement in the $x_2$ direction, with
lattice spacing determined by the deformation parameter $\theta$. The
operators $\overrightarrow{\sf d}_i$ and $\overleftarrow{\sf d}_i$ are, up
to a factor of the lattice spacing, the corresponding discrete derivatives.
In fact, the transformations (\ref{Deltatheta}) are reminiscent of the
shift operator representation of the corresponding Heisenberg-Weyl group
elements. This property may again be attributed to the mixing of
ultraviolet and infrared scales in noncommutative quantum field theory.
Namely, the local, infinitesimal translations induce finite, discrete
shifts by the scale of noncommutativity, and we recover a continuum
interpretation of the UV/IR mixing that occurs in the non-perturbative
lattice regularization of noncommutative field theories~\cite{AMNS2}. The
unusual nature of these point symmetries is a consequence of the inherent
non-locality of the gauge theory.

A similar property holds for the Poincar\'e algebra. The extra generators are
given by
\bea
K_{12}&=&\frac1\theta\,T_{\vec e_2} \ ,
\nn\\ P_1&=&\sum_{n=0}^\infty\frac1{n!}\,T_{n\vec e_1} \ ,
\nn\\ P_2&=&\sum_{n=0}^\infty\frac{(-1)^n}{n!}\,T_{n\vec e_1} \ ,
\label{Poincareundef}\eea
and they realize the undeformed Poincar\'e algebra $iso(1,1)$. Here the
``boost'' $K_{12}$ acts by translations in the $x_1$ direction, so that the
lines of constant $x_1$ may be interpreted as different inertial reference
frames. The space and time translations $P_1$ and $P_2$ correspond to
discrete shifts by the noncommutativity scale along the backward and
forward light-cone directions, respectively.

Evidently there are many other examples that can be constructed along
similar lines, in which a set of transformations has the algebraic
structure of a known transformation group of the spacetime (or some other
Lie group). The general feature will always be the same. Either the
operators act canonically on the spacetime coordinates but yield
deformations of the commutation relations of the Lie algebra, or else they
realize the Lie algebra exactly but display highly non-local effects that
make their geometrical interpretations differ enormously from the
anticipated ones. We shall see in the next section how these unusual
effects come up again in the gauge transformations of the fields. In all of
these cases the exotic behaviours are the characteristic properties of the
spacetime symmetries induced by noncommutative gauge theories. Indeed, the
properties unveiled here illustrate that it is not entirely correct to
regard a noncommutative gauge theory as an ordinary one on a commutative
space in which a noncommutative algebra is defined.

\subsection{Infinite-dimensional Algebras}

We will now describe some simple infinite-dimensional subalgebras of
$u(\alg_\theta)$. First of all, there are $d$ obvious abelian subalgebras,
generated by $T_{n\vec e_i}$, $n\in\zed_+$, for each $i=1,\dots,d$. These
algebras yield gauge transformations by functions which depend only on a
single coordinate $x_i$ and correspond to a choice of ``configuration''
subspace of noncommutative $\real^d$. However, such particular choices are
merely a matter of a choice of axes in $\real^d$. More generally, changing
basis we find that the linear combinations of the form
\be
C_n(\vec c\,)=\sum_{i=1}^dc_i\,T_{n\vec e_i} \ , ~~ n\in\zed_+ \ ,
\ee
generate an infinite-dimensional abelian subalgebra of $u(\alg_\theta)$ for
all {\em constant} real-valued $d$-vectors $\vec c$. Thus there actually
exists a continuous $d$-parameter family of such sorts of abelian
subalgebras.

Formally, abelian subalgebras are parametrized by a Lagrangian submanifold
of the underlying symplectic space, i.e. a subspace on which the symplectic
form vanishes, and correspond to a foliation into symplectic leaves on each
of which the Casimir one-form of the Lie algebra $u(\alg_\theta)$ vanishes.
A more non-trivial choice of Lagrangian submanifold corresponds to the
$2^{d/2}$ classes of ``diagonal'' subalgebras of $u(\alg_\theta)$. In each
commuting two-dimensional skew block $a$, these are generated by the
operators $T_{\underline{n}}^{(a)}$, $n\in\zed_+$, which according to
(\ref{commrs}) have the property
\be
\left[T_{\underline{n}}^{(a)}\,,\,T_{\underline{m}}^{(a)}
\right]^{~}_\star=0 \ .
\label{diagonal}\ee
Closely related to this subalgebra is the one generated by the odd Wigner
functions $\Sigma^{(a)-}_{\underline{n}}$, $n\in\zed_+$, in the density
matrix basis. It consists of radially symmetric functions, i.e. those which
depend only on the real variables $|z_a|$, and it is generated by the
completion of the Cartan subalgebra of the infinite unitary Lie algebra
$u(\infty)$.
Although similar, these two abelian algebras are not the same, for example
it is easy to check by using (\ref{explbasischange}) that their generators do
not mutually commute.

These abelian subalgebras can each be regarded as the local symmetry
algebra of a {\em commutative} $\frac d2$-dimensional $U(1)$ gauge theory.
This opens up the possibility of a mechanism of dimensional reduction for
which the gauge theory on a $d$-dimensional noncommutative space has a
vacuum which is invariant only under the gauge transformations
corresponding to an ordinary gauge theory on a $\frac d2$-dimensional
commutative subspace. The target space will, however, manifest its true
nature at energies of order $\theta^{-1/2}$ whereby, in the string picture,
the effects of the background $B$-field become important. Considering the
complete reduction of the noncommutative dimensions may lead to a sort of
universal gauge theory, containing all Yang-Mills theories, along the lines
described in~\cite{Rajeev}. We shall describe some consequences of this
simple observation in the next section.

Finally, let us point out that there is a set of $d$ non-abelian
infinite-dimensional subalgebras generated by the $T_{\vec n}$'s with a
particular component integer $n^i$ of odd parity. This follows trivially
from the fact that the sum of three odd integers ($n^i$, $m^i$ and $2p+1$
in \eqn{commrs}) is odd. The even/odd parity of $n^i$ for each
$i=1,\dots,d$ gives an extra $\zed_2^d$ grading to the $\star$-monomial
representation of $u(\alg_\theta)$ which is related to the $\pm$ grading of
the density matrix basis as described in section~3.3. It would be
interesting to construct a genuine $W_{1+\infty}$ subalgebra of
$u(\alg_\theta)$ in this setting, which at present only appears as the
approximate symmetry algebra (\ref{W1infty}) at order $\theta_{ij}$.

\newsection{Applications}

In this section we will briefly describe some implications of the analysis
of the previous sections by applying the formalism to study some of the
physical characteristics of noncommutative gauge theories. First we will
discuss some further aspects of the relationships between inner
automorphisms, gauge transformations and spacetime diffeomorphisms. Then we
will describe some aspects of the geometrical structure of solitons in
noncommutative gauge theory.

\subsection{Geometry of Noncommutative Gauge Transformations}

For the bulk of this paper thus far, we have focused on the geometrical
interpretation of inner automorphisms of the algebra $\alg_\theta$, without
saying what the implications are for the corresponding inhomogeneous gauge
transformations in (\ref{ncgaugetransf}). We are now ready to lend some
geometrical insights into the nature of the local $\star$-gauge symmetry of
noncommutative Yang-Mills theory. For this, we expand a generic noncommutative
gauge field $A_i(x)$ over $\complex[[x_1,\dots,x_d]]/{\cal R}_\theta$ in terms
of the momentum operators (\ref{Piops}) as~\cite{munich}
\beq
A_i\equiv-\sum_{j=1}^d\theta_i^{~j}\,T_{\vec e_j}+\Pi_i \ .
\label{Xidef}\eeq
An essential feature of gauge theory on a noncommutative space is the gauge
invariance of the spacetime coordinates up to a transformation of the form
\beq
\delta_\lambda x_i=c_i+\Lambda_i^{~j}\,x_j \ ,
\label{xigaugeinv}\eeq
where $c_i$ are constants parametrizing the (trivial) center $\complex$ of the
algebra, and $\Lambda\in Sp(d)$ are constant matrices which parametrize the
group of rotations that leaves the noncommutativity parameters invariant, i.e.
\beq
\Lambda_i^{~k}\,\theta_{kl}\,\Lambda_j^{~l}=\theta_{ij} \ .
\label{Lambdatheta}\eeq
Only in that case is the algebra (\ref{commx},\ref{alggenxi}) preserved by the
gauge transformation. Unless specified otherwise, we shall take
$c_i=\Lambda_i^{~j}=0$ in (\ref{xigaugeinv}) for simplicity.

The fields $\Pi_i$ which are defined by (\ref{Xidef}) may be thought of as
gauge covariant momentum operators, because the gauge transformation rule
(\ref{ncgaugetransf}) is equivalent to the infinitesimal inner automorphism
\beq
\delta_\lambda\Pi_i=i\,[\lambda,\Pi_i]^{~}_\star \ .
\label{Xicovtransf}\eeq
Moreover, the entire noncommutative gauge theory can be expressed in terms of
the new fields $\Pi_i$. The gauge covariant derivative (\ref{nablaadjoint}) may
be written as the inner derivation
\beq
\nabla_i^{\rm ad}f=i\,[f,\Pi_i]^{~}_\star \ , ~~ \forall f\in\alg_\theta \ ,
\label{nablaadXi}\eeq
while the field strength tensor (\ref{NCfieldstrength}) is the sum
\beq
F_{ij}=-i\,[\Pi_i,\Pi_j]^{~}_\star+B_{ij} \ .
\label{FijXi}\eeq
The classical vacua of the noncommutative Yang-Mills theory (\ref{NCYM}),
i.e. the flat noncommutative gauge connections, are in this setting the
noncommuting covariant momenta with $[\Pi_i,\Pi_j]_{\star}^{~}=-i\,B_{ij}$. In
the D-brane picture, the corresponding global minima are identified with the
closed string vacuum possessing no open string excitations. The transition from
the momentum operators $P_i$ to the covariant ones $\Pi_i$ reflects the
bi-module structure based on the noncommutative function algebras with
deformation parameters $\theta$ and $-\theta$, and it is the basis of the
relation between the commutative and noncommutative descriptions of the same
theory~\cite{CNCrel}. This gives a dynamical interpretation to the
noncommutativity of spacetime.

Thus the local noncommutative gauge symmetry is determined entirely by the
inner automorphisms (\ref{Xicovtransf}), and in this way the geometrical
interpretations given in the previous section carry through to the
noncommutative gauge fields. Let us now examine these transformations in
some detail. The simplest ones are the spacetime translations generated by
the operators (\ref{Piops}). Notice first that although the functions
(\ref{Piops}) only form a projective representation of the $d$-dimensional
translation group, the corresponding Lie algebra (\ref{deltacomm}) of gauge
transformations forms a true representation,
\beq
\left[\delta_{P_i}\,,\,\delta_{P_j}\right]=0 \ ,
\label{deltatruerep}\eeq
because the projective phase in (\ref{NCtranslalg}) lies in the center
$\complex$ of the algebra $\alg_\theta$. The emergence of this true
representation of the translation group owes to the effective enlargement
of the deformed group representation via the decomposition (\ref{Xidef}).
Namely, the gauge transform (\ref{ncgaugetransf}) in this case reads
\beq
\delta_{P_j}A_i=\partial_jA_i+B_{ij} \ ,
\label{deltaPjAi}\eeq
which up to the addition of the magnetic flux $B_{ij}$ is the anticipated
transformation rule for a one-form field under spacetime translations.
Again (\ref{deltaPjAi}) corresponds to a projective representation of the
translation group. This is yet another manifestation of UV/IR mixing, in
that a translation along a direction $j$ causes a constant shift of the
gauge fields $A_i$ along the orthogonal noncommutative directions $i$.
Constant shifts $A_i\mapsto A_i+c_i$ of the gauge fields correspond to
global gauge transformations by the center of the algebra, and are
therefore global symmetries of the field theory (\ref{NCYM}). Indeed, the
field strength tensors (\ref{NCfieldstrength}) are invariant under such
shifts. The fact that a local gauge transformation induces a global
symmetry is again due to the non-locality of the noncommutative gauge
theory.

Let us now consider the rotation generators (\ref{L12def}). We immediately
encounter two important differences from the case above. First, the
inhomogeneous terms in (\ref{ncgaugetransf}) that appear in the
noncommutative directions are the functions $\theta^{-1}\,T_{\vec e_i}$,
which are non-central elements of the algebra. This owes to the more
complicated group composition law (\ref{quadgrouplaw},\ref{Lambdaeff}) for
Gaussian functions, or equivalently to the non-central extensions of the
deformed $iso(d)$ algebras (\ref{isodgen}). Second, the corresponding inner
automorphism (\ref{Xicovtransf}) generates a rotation of a {\it scalar} field,
not a one-form field. In other words, the noncommutative gauge symmetry ignores
the vector index of the noncommutative gauge fields. The reason for this will
be explained
below. Analogous but more involved properties are true of the generators
(\ref{E2undef}). In that case, the corresponding gauge transformations
(\ref{undefinner}) (along with the deformed translational symmetry)
immediately imply the inherent non-locality of noncommutative Yang-Mills
theory, i.e. that gauge invariant observables are necessarily non-local.
These structures all come about from the mixing of colour and spacetime
symmetries that we have alluded to earlier.

By using the covariant momentum operators we may also examine to what extent a
generic local diffeomorphism of $\real^d$ can be realized as a genuine gauge
symmetry of the noncommutative Yang-Mills theory. Given an arbitrary, local
vector field $V=V^i(x)\,\partial_i$ on $\real^d$, we introduce a corresponding
gauge function $\lambda=\lambda_V$ (again over $\complex[[x_1,\dots,x_d]]/{\cal
R}_\theta$) by the $\star$-anticommutator
\beq
\lambda_V=-\frac12\,\sum_{j=1}^d\theta_i^{~j}\,\left\{T_{\vec e_j}\,,\,V^i
\right\}^{~}_\star \ .
\label{lambdaV}\eeq
Then the corresponding gauge transformation (\ref{Xicovtransf}) can be expanded
in the deformation parameters by using (\ref{starcomm}) and
(\ref{staranticomm}) to obtain the leading order result
\beq
\delta_{\lambda_V}\Pi_i=V^j\,\partial_j\Pi_i+\sum_{j=1}^d\theta_k^{~j}\,
T_{\vec e_j}\,\delta^{lm}\,\delta^{np}\,\theta_{mp}\,\partial_lV^k\,
\partial_n\Pi_i+O(\theta_{ij}) \ .
\label{deltalambdaVPi}\eeq
The first term in (\ref{deltalambdaVPi}) is close to the expected
transformation law for $\Pi_i$ under an infinitesimal diffeomorphism, except
that it treats it as a scalar field. This is not surprising, because the gauge
theory we have formulated is defined on a flat space and so only possesses a
{\it global} Lorentz symmetry, and not a local one. In other words, only those
diffeomorphisms which are isometries of flat spacetime arise in noncommutative
gauge transformations. This feature is described further in~\cite{LSred}, and
possible extensions of noncommutative Yang-Mills theory to incorporate local
frame independence are analysed in~\cite{LSred,AIKO}. In any case, this term is
accompanied by the second term in (\ref{deltalambdaVPi}) which is of the same
order in $\theta_{ij}$. This fact on its own prevents one from realizing
arbitrary diffeomorphisms in terms of $\star$-gauge symmetries. One could
nonetheless carry on and attempt to interpret the first term in
(\ref{deltalambdaVPi}) as a vierbein transformation rule for a flat space
vierbein field $h_i^j$, as in~\cite{AIKO}, defined through the decomposition
$\Pi_i=\sum_j\theta_k^{~j}\,T_{\vec e_j}\,h_i^k$. But this is only possible
when the spacetime coordinates themselves gauge transform as
$\delta_{\lambda_V}x_i=V_i(x)$. Unless the vector field $V$ is parametrized by
an element of the Lie algebra $\complex\oplus sp(d)$ as in (\ref{xigaugeinv}),
such a transformation will map noncommutative $\real^d$ onto a different
noncommutative space and will not be a symmetry of the theory. The resultant
space would be related to the dynamics of D-branes in non-constant $B$-fields,
which are described in terms of non-associative algebras~\cite{CorSch}. Such a
manipulation will thereby only work for the Gaussian spacetime transformations
that we described in section~5.1. Again this is just an indication of the basic
fact that only (deformations of) symplectomorphisms are realized as
noncommutative gauge transformations.

\subsection{Conformal Non-invariance}

In previous sections we have seen how gauge transformations are given as inner
automorphisms of the algebra $\alg_\theta$ and that they correspond to
symplectic diffeomorphisms. We can also use the present formalism to
investigate this geometrical feature in a bit more detail. Not all inner
automorphisms of $\alg_\theta$ correspond to gauge transformations, but only
the ones which are generated by Hermitian, Schwartzian elements. It is also
straightforward to show explicitly that there exist diffeomorphisms of
$\real^d$ which cannot be realized via $\star$-commutators. For example, let us
consider the scale transformation $x_i\mapsto x_i+\alpha_i\,x_i$, $i=1,\dots,d$
(no sum on $i$), with $\alpha_i$ real-valued constants. To realize this
coordinate transformation as an infinitesimal inner automorphism of
$\alg_\theta$, we seek an element $S_\alpha$ of the algebra with the properties
\bea
S_\alpha&=&\sum_{\vec m\in\zeds_+^d}s_{\vec m}\,T_{\vec m} \ , \non
i\,[S_\alpha,T_{\vec n}]^{~}_\star&=&\sum_{i=1}^d\alpha_i\,n^i\,T_{\vec n}
\ , ~~ \forall\vec n\in\zed_+^d\ .
\label{scaleansatz}\eea
Taking $\vec n=\vec e_1$ in (\ref{scaleansatz}), and using (\ref{cnmp}),
gives $s_{\underline{1}}=\alpha_1$ and $s_{\vec m}=0~~\forall\vec
m\neq\underline{1}$. On the other hand, taking $\vec n=\vec e_2$ in
(\ref{scaleansatz}) gives $s_{\underline{1}}=-\alpha_2$, which generically
leads to a contradiction. This proves that conformal transformations cannot
be obtained via inner automorphisms of $\alg_\theta$ and do not constitute
gauge symmetries of noncommutative Yang-Mills theory. Notice that this
argument is independent of any reality assumption or asymptotic behaviour of
the expansion coefficients $s_{\vec m}$ of (\ref{scaleansatz}).

The lack of a conformal gauge symmetry can also be seen algebraically as
follows. In the skew-diagonalization, the truncated Lie algebra defined by
the first line of \eqn{truncated} also contains, in each skew-block, an
infinite-dimensional subalgebra which corresponds to the positive Borel
subalgebra of the classical Virasoro-Witt algebra. By defining
\bea
\ell_n&=&\frac 1\theta\,T_{n\vec e_1+\underline{1}} \ , ~~ n\geq0 \ , \nn\\
\ell_{-1}&=&\frac1\theta\,T_{\vec e_2}
\label{halfvir}\eea
in the two-dimensional case, we arrive at the $\star$-commutation relations
\be
[\ell_n,\ell_m]^{~}_\star=i\,(n-m)\,\ell_{n+m} \ .
\label{ellstarcomm}\ee
This algebra contains an $su(2)$ Lie subalgebra generated by the functions
$\ell_0$ and $\ell_{\pm1}$. It is possible to generate a full Virasoro-Witt
algebra by defining instead the operators
\bea
L_n=\left\{\new{\begin{array}{l}\frac1\theta\,\left(T_{n\vec e_1+
\underline{1}}+ \frac{i\,\theta}{2~n!}\,T_{n\vec e_1}\right) \ , ~~ n\geq0 \ ,
\\\frac1\theta\,T_{\vec e_2} \ , ~~ n=-1 \ , \\
\frac1\theta\,\sum_{k=0}^{\infty}~\sum_{l=0}^k\left(\begin{array}{c}k\\l
\end{array}\right)\,\Biggl(T_{(|n|-1)(k-l)\vec e_1+\underline{1}}\Biggr.\\+
\left.\frac{i\,\theta}{2~\Bigl((|n|-1)(k-l)\Bigr)!}\,T_{\bigl((|n|-1)(k-l)-1
\bigr)\vec e_1}\right) \ , ~~ n<-1 \ . \end{array}} \right.
\label{virseries}\eea
The third expression in (\ref{virseries}) is to be understood as a formal
series, because it strictly speaking only makes sense, as a Weyl operator say,
on the domain of Hilbert space in which the eigenvalues of the Hermitian
operator $\hat x_1$ lie in the subset $\real_-\cup[2^{1/n},\infty)$ of the real
line. Then, by using (\ref{commrs}) we may infer the $\star$-commutation
relations
\beq
[L_n,L_m]^{~}_\star=i\,(n-m)\,L_{n+m} \ .
\label{fullviralg}\eeq
Interchanging the unit lattice vectors $\vec e_1$ and $\vec e_2$ in
(\ref{virseries}) defines corresponding ``anti-holomorphic'' generators
$\overline{L}_n$. It may be straightforwardly checked that the operators
(\ref{virseries}) (along with the $\overline{L}_n$) generate the anticipated
conformal transformations of scalar fields in two dimensions via infinitesimal
inner automorphisms. However, although we have obtained the correct commutation
relations of the conformal algebra in two dimensions, the realization
(\ref{virseries}) defines a non-unitary representation of this Lie algebra in
the Moyal bracket. In addition, $[L_n,\overline{L}_m]^{~}_\star\neq0$, so that
the two copies of the algebra are not independent. This demonstrates again that
there is no unitary realization of conformal transformations as noncommutative
gauge symmetries on the Moyal space.

\subsection{Geometrical Structure of Noncommutative Solitons}

A concrete realization of the breaking of the $u(\alg_\theta)$ gauge symmetry
to the infinite-dimensional commutative subalgebras described in section~5.3 is
provided by the soliton configurations of noncommutative field
theory~\cite{NCsolitons}. They are determined by projection operators or
partial isometries in the unital algebra
$\alg^+_\theta=\alg_\theta\oplus\complex$. For simplicity, we shall again
restrict to the two-dimensional case. For $U(N)$ noncommutative gauge theory,
the static soliton solutions are given by the covariant momentum
operators~\cite{GN,Bak}
\beq
\Pi_i={\sf S}\,T_{\vec e_i}\,{\sf S}^\dagger+\sum_{\mu=1}^m\alpha_i^\mu\,
\Sigma_{\underline{\mu}}^-
\label{Piisoliton}\eeq
for $i=1,2$, where the density matrices $\Sigma^-_{\underline{\mu}}$,
$\mu=1,\dots,m$, correspond to an $m$-dimensional subspace of the Hilbert space
$\ell^2(\zed_+)\otimes\complex^N$, and $\sf S$ is the associated shift operator
which is the partial isometry in $\mat_N(\alg_\theta^+)$ obeying
\beq
{\sf S}^\dagger\star{\sf S}=\id \ , ~~ {\sf S}\star{\sf S}^\dagger=
\id-\sum_{\mu=1}^m\Sigma^-_{\underline{\mu}} \ .
\label{shiftop}\eeq
The $2m$ moduli parameters $\alpha_i^\mu$ of the classical gauge field
configuration (\ref{Piisoliton}) may be interpreted as the locations of
D0-branes (the solitons) inside D2-branes (the noncommutative transverse
space), with $m$ and $N$ the 0-brane and 2-brane charges, respectively.

To understand better the soliton solutions (\ref{Piisoliton}) within the
present context, we will study the somewhat simpler case of static solitons in
a noncommutative {\it scalar} field theory in the limit of large
noncommutativity $\theta$. As is well-known~\cite{NCsolitons}, the space of
such solutions is spanned, for a given scalar potential, by an orthonormal
basis of projectors ${\sf P}_n$, $n\in\zed_+$, in the algebra $\alg_\theta^+$,
\bea
{\sf P}_n\star{\sf P}_m&=&\delta_{nm}\,{\sf P}_n \ , \nn\\\sum_{n=0}^\infty
{\sf P}_n&=&1 \ .
\label{projectors}\eea
Using the Weyl-Wigner correspondence (\ref{WWcorr}) and the density matrix
basis (\ref{taualphahatdef}), it is easy to see that a basis of functions
satisfying (\ref{projectors}) is given by
\beq
{\sf P}_n=\mbox{$\frac12$}\,\Sigma_{\underline{n}}^- \ .
\label{projSigma}\eeq
By using (\ref{starunitarity}), we see that the symmetry group of the equations
(\ref{projectors}) is precisely the unitary group $U(\alg_\theta^+)$. Each
non-vanishing solution ${\sf P}_n(|z|)$ breaks this symmetry down to a
commutative subgroup. A rank $m$ projection operator will induce a broken
$U(m)$ subgroup of $U(\alg_\theta^+)$ corresponding to the symmetry group of
$m$ coincident D-branes. Inner automorphisms of the algebra $\alg_\theta^+$
rotate the basis (\ref{projSigma}) to generically non-radially symmetric
soliton solutions. For $\theta<\infty$, the kinetic term for the scalar field
will explicitly break the unitary gauge symmetry down to the Euclidean subgroup
$ISO(2)$ of $U(\alg_\theta^+)$. This lifts the manifold of soliton solutions to
a discrete set of solutions.

The non-trivial structure of the gauge group can be seen explicitly in this
context by approximating these soliton solutions by the finite-rank operators
constructed in section~4. Given the Cartan and Weyl bases (\ref{hatEdef}) and
(\ref{hatJNdef}), we may use a discrete version of the Wigner transform
(\ref{Wigner}) to define the functions
\beq
E_{\vec n}(x)=\frac1N\,\sum_{\vec m\in\zeds_N^2}\e^{2\pi i\,\vec m\cdot x/N}~
\Tr_N\left(\hat E_{\vec n}\,\hat{\cal W}_{\vec m}^{(N)}\right) \ ,
\label{Wignerdiscrete}\eeq
where $\vec n\in\zed_N^2$ and here $x_i$ are interpreted as coordinates on a
periodic, square lattice of spacing $\frac1N$. This corresponds to a
finite-dimensional representation of the Heisenberg-Weyl group. By using the
change of basis (\ref{hatEJchange}), the orthonormality relations
(\ref{hatJortho}), and the lattice completeness relations
\beq
\frac1N\,\sum_{\vec m\in\zeds_N^2}\e^{2\pi i\,\vec m\cdot(x-y)/N}=N^2\,
\delta(x-y) \ ,
\label{complreldiscrete}\eeq
the Wigner functions (\ref{Wignerdiscrete}) can be written as
\beq
E_{\vec n}(x)=-4\pi MN\,\omega^{(n^1-n^2)(n^1-2n^2-1)/2}~\e^{-2\pi i\,
(n^1-n^2)x_2/N}\,\delta\Bigl(x_1+M(n^1-2n^2)\Bigr) \ .
\label{WignerEfinal}\eeq
We may thereby approximate the solutions of the noncommutative soliton
equations (\ref{projectors}) by the semi-localized finite-rank projection
operators
\beq
{\sf P}_n^{(N)}(x)=E_{\underline{n}}(x)=-4\pi MN\,\delta(x_1-Mn) \ .
\label{projfiniteN}\eeq
Such fuzzy soliton configurations have also been obtained
in~\cite{fuzzysolitons}. Note that in this lattice basis the coordinate $x_1$
plays a role analogous to the continuum radial coordinate $|z|$. This is
expected from the properties of the rotation generators described in
section~5.2, namely that the true representation of the Euclidean group will
affect rotations in the $x_1$ coordinate, and hence of the soliton.

The discrete projectors (\ref{projfiniteN}) illustrate the non-trivial moduli
space of soliton solutions that arise, which are parametrized by their location
$x_1$. They describe stripes on the Moyal plane labelled by the integers $M$
and $n$. However, they do not converge to the continuum projector solutions in
the large $N$ limit, which are given in terms of Laguerre polynomials as in
section~3. This illustrates the basic importance of the large $N$ completion in
(\ref{gaugesdiff}) that defines the unitary group of the algebra
$\alg_\theta^+$, and it can be understood in terms of its corresponding
K-theory~\cite{Weggeolsen}. While the group $\hk_0(\alg_\theta^+)$ is given in
terms of partial-unitary equivalence classes of projectors in
$\alg_\theta^+\otimes\mat_\infty(\complex)$, the $\hk_1$-group is given in
terms of the connected components of the gauge group as
\beq
\hk_1\left(\alg_\theta^+\right)=U\left(\alg_\theta^+\right)\,/\,U
\left(\alg_\theta^+\right)^{~}_0 \ ,
\label{K1alg}\eeq
where $U(\alg_\theta^+)^{~}_0$ denotes the (local) connected subgroup of
$U(\alg_\theta^+)$. With this definition it is evident that the $\hk_1$-group
of any finite-dimensional matrix algebra $\mat_N(\complex)$ is trivial, and,
since K-theory is stable under inductive limits, so is that of
$\mat_\infty(\complex)$ as defined in section~4.2. On the other hand, the group
(\ref{K1alg}) is known to be non-trivial in certain instances. Therefore, an
appropriate completion of the infinite unitary group $U(\infty)$ is required to
preserve the K-theoretic properties of the Moyal space. Such completions in the
case of the noncommutative torus are described in~\cite{largeNnc}.

Of course, the applications described in this subsection require an analysis of
the {\it global} gauge group, as in (\ref{K1alg}), which lies beyond the scope
of this paper. They are important however for many of the applications of these
results to the description of D-branes as solitons in the effective field
theory of open strings. Indeed, D-brane charges are classified by the K-theory
groups of the spaces on which they are defined~\cite{KTheory,WittenK}. They
correspond to projectors or partial isometries arising as solitonic lumps of
the tachyon field in a given non-BPS system of higher-dimensional
D-branes~\cite{NCsolitons}. The mechanisms for symmetry breaking described in
section~5.3 can thereby lend a more geometrical picture to processes involving
tachyon condensation, and, in particular, the low-energy effective string field
theory action~\cite{WittenNC} may in this setting correspond to some sort of
universal gauge theory~\cite{Rajeev}. In this context, the automorphism group
(\ref{Autcompactgen}) corresponds to the noncommutative gauge group of D-branes
in the presence of NS5-branes~\cite{NCtachyons}, i.e. in a topologically
non-trivial $B$-field, which may be realized as certain twisted $PU(\hil)$
bundles~\cite{WittenK,twistK}. The corresponding description of the D-brane
charges is known to be given by a twisted version of
K-theory~\cite{WittenK,twistK}.

\newsection{Conclusions}

The intriguing mixing between the infrared and ultraviolet limits is a
characteristic feature of gauge theories on noncommutative spaces, and it is
intimately tied to the mixing between spacetime variables and gauge degrees of
freedom. This is reflected in the gauge transformations of the fields which
drastically alter not only their internal degrees of freedom, but also their
spacetime dependence. It is a consequence of the fact that, strictly speaking,
it does not makes sense to speak of ``point dependence'' for these field
theories, as the concept of a point is ill-defined. In particular, it is not
possible to regard the gauge algebra as the tensor product of a finite
dimensional algebra by the set of functions on a point. This implies that, like
in general relativity, the spacetime structure is a gauge non-invariant
concept. We have seen that the gauge degrees of freedom come from a deformation
of the Poisson-Lie algebra of symplectic diffeomorphisms. These gauge
transformations are intimately related to unitary conjugation by elements of
the group $U(\infty)$, which is the natural symmetry group that arises from the
large $N$ matrix models which provide non-perturbative regularizations of
noncommutative gauge theory.

It is tempting at this point to speculate on the relationships between these
deformed canonical transformations and time evolution. A dynamical system can
be thought of as a $C^*$-algebra together with a one-parameter group of
symplectic automorphisms which is generated by a Hamiltonian. In the
noncommutative case, gauge transformations are equivalent to canonical
transformations of spacetime. It would be interesting to interpret this
deformed $U(\infty)$ gauge symmetry in terms of the group of Hamiltonian flows.

The sort of noncommutative geometry discussed in this paper is of course but an
approximation to the full string theory. The general structure of spacetime at
the string (or Planck) scale is likely to be described by the $*$-algebra
$\alg_{\rm str}$ of open string fields which is defined by the gluing together
of strings. In the low-energy scaling limit under consideration, this algebra
factorizes as~\cite{WittenNC} $\alg_{\rm str}\to\alg_{\rm
v}^0\otimes\alg_\theta$, where $\alg_{\rm v}^0$ is the algebra of vertex
operators with vanishing momentum along the noncommutative directions, and
$\alg_\theta$ is the noncommutative function algebra considered in this paper.
If the deformed space we have dealt with is an accurate description of
spacetime at energies of the order $\theta^{-1/2}$, then at these energies
gauge invariance transforms (via deformed symplectic diffeomorphisms) the very
structure of spacetime. Given that, in the noncommutative geometry based on
closed string vertex operator algebras~\cite{LLS}, generic diffeomorphisms of
spacetime can be viewed as gauge transformations, it is possible that the
stringy part of the algebra $\alg_{\rm str}$ reinstates full general covariance
as a genuine gauge symmetry. In the low-energy limit above, a spacetime
diffeomorphism then has a ``gauge part'' related to the spacetime algebra
$\alg_\theta$ and a ``conformal part'' related to the stringy algebra
$\alg_{\rm v}^0$.

It is also possible that the effective string field theory action will only be
invariant under a subset of the possible gauge transformations, leading to
theories for which the accessible gauge theory is much smaller. The resultant
gauge symmetry may either still be infinite-dimensional, corresponding to some
sort of dimensional reduction, or it may become finite-dimensional,
corresponding to a total dimensional reduction (along the noncommutative
directions) with induced internal degrees of freedom. In this respect we could
see the emergence of $u(N)\subset u(\infty)$ gauge models, which are otherwise
difficult to obtain in gauge theories based on noncommutative geometry.

\subsection*{Acknowledgments}

We thank D.~Fairlie, J.~Gracia-Bond\'\i a, E.~Langmann, G.~Marmo,
N.~Mavromatos, M. Schlichenmaier, J.~V\'arilly, P.~Vitale and C.~Zachos for
helpful discussions and correspondence. The work of F.L. was supported in
part by the {\sl Progetto di Ricerca di Interesse Nazionale SINTESI}. The
work of R.J.S. was supported in part by an Advanced Fellowship from the
{\sl Particle Physics and Astronomy Research Council~(U.K.)}.


\begin{thebibliography}{99}

\baselineskip=12pt

\bibitem{book} A.~Connes, {\it Noncommutative Geometry} (Academic Press, 1994).

\bibitem{CDS} A.~Connes, M.R.~Douglas and A.~Schwarz, {\sl J. High Energy
Phys.} {\bf 9802} (1998) 003, {\tt hep-th/9711162}; M.R.~Douglas and C.M.~Hull,
{\sl J. High Energy Phys.} {\bf 9802} (1998) 008, {\tt hep-th/9711165}.

\bibitem{SW} N.~Seiberg and E.~Witten, {\sl J.\ High Energy Phys.} {\bf 9909}
(1999) 032, {\tt hep-th/9908142}.

\bibitem{munich} J.~Madore, S.~Schraml, P.~Schupp and J.~Wess, {\sl Eur. Phys.
J.} {\bf C16} (2000) 161, {\tt hep-th/0001203}.

\bibitem{IIKK} N.~Ishibashi, S.~Iso, H.~Kawai and Y.~Kitazawa, {\sl Nucl.
Phys.} {\bf B573} (2000) 573, {\tt hep-th/9910004}.

\bibitem{AMNS1} J.~Ambj\o rn, Y.M.~Makeenko, J.~Nishimura and R.J.~Szabo, {\sl
J. High Energy Phys.} {\bf 9911} (1999) 029, {\tt hep-th/9911041}.

\bibitem{AMNS2} J.~Ambj\o rn, Y.M.~Makeenko, J.~Nishimura and R.J.~Szabo, {\sl
J. High Energy Phys.} {\bf 0005} (2000) 023, {\tt hep-th/0004147}.

\bibitem{DHI} D.J.~Gross, A.~Hashimoto and N.~Itzhaki, {\sl Adv. Theor. Math.
Phys.} {\bf 4} (2000) No. 4, {\tt hep-th/0008075}; A.~Dhar and S.R. Wadia, {\sl
Phys. Lett.} {\bf B495} (2000) 413, {\tt hep-th/0008144}.

\bibitem{SUGRA} S.R.~Das and S.-J.~Rey, {\sl Nucl. Phys.} {\bf B590} (2000) 453, {\tt hep-th/0008042};
H.~Liu, {\tt hep-th/0011125}; S.R.~Das and S.P.~Trivedi, {\sl
J. High Energy Phys.} {\bf 0102} (2001) 046, {\tt hep-th/0011131}; Y.~Okawa and
H.~Ooguri, {\sl Nucl. Phys.} {\bf B599} (2001) 55, {\tt hep-th/0012218}.

\bibitem{LLS} F.~Lizzi and R.J.~Szabo, {\sl Comm. Math. Phys.} {\bf 197} (1998)
667, {\tt hep-th/9707202}; G.~Landi, F.~Lizzi and R.J.~Szabo, {\sl Comm. Math.
Phys.} {\bf 206} (1999) 603, {\tt hep-th/9806099}.

\bibitem{AIIKKT} H.~Aoki, N.~Ishibashi, S.~Iso, H.~Kawai, Y.~Kitazawa and
T.~Tada, {\sl Nucl. Phys.} {\bf B565} (2000) 176, {\tt hep-th/9908141}.


\bibitem{IKKKK} S.~Iso, H.~Kawai and Y.~Kitazawa, {\sl Nucl. Phys.} {\bf B576}
(2000) 375, {\tt hep-th/0001027}; Y.~Kimura and Y.~Kitazawa, {\sl Nucl. Phys.}
{\bf B598} (2001) 73, {\tt hep-th/0011038}.

\bibitem{FairlieZachos} D.B.~Fairlie, P.~Fletcher and C.K.~Zachos, {\sl Phys.
Lett.} {\bf B218} (1989) 203; {\sl J. Math. Phys.} {\bf 31} (1990) 1088;
D.B.~Fairlie and C.K.~Zachos, {\sl Phys. Lett.} {\bf B224} (1989) 101.

\bibitem{IIKK2} N.~Ishibashi, S.~Iso, H.~Kawai and Y.~Kitazawa, {\sl Nucl.
Phys.} {\bf B583} (2000) 159, {\tt hep-th/0004038}.

\bibitem{LSred} E.~Langmann and R.J.~Szabo, {\tt hep-th/0105094}.

\bibitem{Sochichiu} C.~Sochichiu, {\sl J. High Energy Phys.} {\bf 0005} (2000)
026, {\tt hep-th/0004062}.

\bibitem{AIKO} T.~Azuma, S.~Iso, H.~Kawai and Y.~Ohwashi, {\tt hep-th/0102168}.

\bibitem{Sheikh} M.M.~Sheikh-Jabbari, {\sl J. High Energy Phys.} {\bf 9906}
(1999) 015, {\tt hep-th/9903107}.

\bibitem{NCFTrev} M.R.~Douglas and N.A.~Nekrasov, {\tt hep-th/0106048}.

\bibitem{Harvey} V.P. Nair and A.P. Polychronakos, {\sl Phys. Rev. Lett.}
{\bf 87} (2001) 030403, {\tt hep-th/0102181}; J.A.~Harvey,
{\tt hep-th/0105242}.

\bibitem{CPST1} M.~Chaichian, P.~Pre\v{s}najder, M.M.~Sheikh-Jabbari and
A.~Tureanu, {\tt hep-th/0107037}.

\bibitem{GN} D.J.~Gross and N.A.~Nekrasov, {\sl J. High Energy Phys.} {\bf
0103} (2001) 044, {\tt hep-th/0010090}.

\bibitem{Merk} S.A.~Merkulov, {\tt math-ph/0001039}; Y.~Zunger, {\tt
hep-th/0106030}.

\bibitem{Groenewold} H.~Groenewold, {\sl Physica} {\bf 12} (1946) 405.

\bibitem{Moyal} J.E.~Moyal, {\sl Proc. Cambridge Phil. Soc.} {\bf 45} (1949)
99.

\bibitem{Weggeolsen} N.E.~Wegge-Olsen, {\it K-Theory and $C^*$-algebras}
(Oxford University Press, 1993).

\bibitem{Schwarz} A. Schwarz, {\sl Comm. Math. Phys.} {\bf 221} (2001) 433,
{\tt hep-th/0102182}.

\bibitem{KrieglMichor} A.~Kriegl and P.~Michor, {\it The Convenient Setting of
Global Analysis} (AMS, 1997).

\bibitem{murphy} G.J.~Murphy, {\it $C^*$-algebras and Operator Theory}
(Academic Press, 1990).

\bibitem{ReedSimon} M.~Reed and B.~Simon, {\it Functional Analysis} (Academic
Press, 1972).

\bibitem{BLP} A.P. Polychronakos, {\sl J. High Energy Phys.} {\bf 0011} (2000)
008, {\tt hep-th/0010264};
D.~Bak, K.~Lee and J.-H.~Park, {\sl Phys. Lett.} {\bf B501}
(2001) 305, {\tt hep-th/0011244}.

\bibitem{realncg} A.~Connes, {\sl J.\ Math.\  Phys.} {\bf 36} (1995) 6194; {\sl
Comm. Math. Phys.} {\bf 155} (1996) 109, {\tt hep-th/9603053}.

\bibitem{Weyl} H.~Weyl, {\it The Theory of Groups and Quantum Mechanics}
(Dover, 1931).

\bibitem{Wigner} E.P.~Wigner, {\sl Phys.\ Rev.} {\bf 40} (1932) 749.

\bibitem{CUZ} T.~Curtright, T.~Uematsu and C.K.~Zachos, {\sl J. Math. Phys.}
{\bf 42} (2001) 2396, {\tt hep-th/0011137}.

\bibitem{Gradshteyn} I.S.~Gradshteyn and I.M.~Ryzhik,
{\it Table of Integrals, Series, and Products} (Academic Press, 1980).

\bibitem{Jgauge} L.~Bonora, M.~Schnabl, M.M.~Sheikh-Jabbari and A.~Tomasiello,
{\sl Nucl. Phys.} {\bf B589} (2000) 461, {\tt hep-th/0006091}; B.~Jur\v{c}o,
S.~Schraml, P.~Schupp and J.~Wess, {\sl Eur. Phys. J.} {\bf C17} (2000) 521,
{\tt hep-th/0006246}; I.~Bars, M.M.~Sheikh-Jabbari and M.A.~Vasiliev, {\tt
hep-th/0103209}.

\bibitem{volpres} Y.~Matsuo and Y.~Shibasa, {\sl J. High Energy Phys.} {\bf
0102} (2001) 006, {\tt hep-th/0010040}.

\bibitem{GozziReuter} E.~Gozzi and M.~Reuter, {\sl Int.\ J.\ Mod.\ Phys.} {\bf
A9} (1994) 5801.

\bibitem{BFFLS} F.~Bayen, M.~Flato, C.~Fronsdal, A.~Lichnerowicz and
D.~Sternheimer, {\sl Ann. Phys.} {\bf 111} (1978) 61.

\bibitem{UVIR} S.~Minwalla, M.~Van~Raamsdonk and N.~Seiberg, {\sl J. High
Energy Phys.} {\bf 0002} (2000) 020, {\tt hep-th/9912072}.

\bibitem{Rieffel} M.A.~Rieffel, {\sl Duke Math.\ J.} {\bf 39} (1972) 745.

\bibitem{AgarwalWolf}  J.C.~Pool, {\sl J. Math. Phys.} {\bf 7} (1966) 66;
G.S.~Agarwal and E.~Wolf, {\sl Phys. Rev.} {\bf D2} (1970) 2161.

\bibitem{Howe} R.~Howe, {\sl J. Funct. Anal.} {\bf 38} (1980) 188.

\bibitem{Ticos} J.M.~Gracia-Bond\'\i a, J.C.~V\'arilly and H.~Figueroa, {\it
Elements of Noncommutative Geometry} (Birkh\"auser, 2000).

\bibitem{completion} A.~Voros, {\sl J.\ Funct.\ Anal.} {\bf 41} (1978) 104;
I.~Daubechies, {\sl Comm.\ Math.\ Phys.} {\bf 75} (1980) 229; {\sl J.\ Math.\
Phys.} {\bf 24} (1983) 1453; J.C.~V\'arilly and J.M.~Gracia-Bond\'\i a, {\sl
J.\ Math.\ Phys.} {\bf 29} (1988) 880.

\bibitem{Kac} V.G.~Kac, {\it Infinite-dimensional Lie Algebras} (Cambridge
University Press, 1985).

\bibitem{HoppeSchaller} J.~Hoppe and P.~Schaller, {\sl Phys.\ Lett.} {\bf B237}
(1990) 407.

\bibitem{largeNnc} G.~Landi, F.~Lizzi and R.J.~Szabo, {\sl Comm. Math. Phys.}
{\bf 217} (2001) 181, {\tt hep-th/9912130}.

\bibitem{BHSS} M.~Bordemann, J.~Hoppe, P.~Schaller and M.~Schlichenmaier, {\sl
Comm.\ Math.\ Phys.} {\bf 138} (1991) 207.

\bibitem{BFSS} T.~Banks, W.~Fischler, S.H.~Shenker and L.~Susskind, {\sl Phys.
Rev.} {\bf D55} (1997) 5112, {\tt hep-th/9610043}.

\bibitem{dWHN} B.~de~Wit, J.~Hoppe and H.~Nicolai, {\sl Nucl. Phys.} {\bf B305}
(1988) 545.

\bibitem{IKKT} N.~Ishibashi, H.~Kawai, Y.~Kitazawa and A.~Tsuchiya, {\sl Nucl.
Phys.} {\bf B498} (1997) 467, {\tt hep-th/9612115}.

\bibitem{CNCrel} L.~Cornalba, {\sl Adv. Theor. Math. Phys.} {\bf 4} (2000) 271,
{\tt hep-th/9909081}; N.~Ishibashi, {\tt hep-th/9909176}; B.~Jur\v{c}o and
P.~Schupp, {\sl Eur. Phys. J.} {\bf C14} (2000) 367, {\tt hep-th/0001032}.

\bibitem{HKL} J.A. Harvey, P. Kraus and F. Larsen, {\sl J. High Energy Phys.}
{\bf 0012} (2000) 024, {\tt hep-th/0010060}.

\bibitem{Lichner} A.~Lichnerowicz, {\sl J. Diff. Geom.} {\bf 12} (1977) 253.

\bibitem{NCOS} N.~Seiberg, L.~Susskind and N.~Toumbas, {\sl J. High Energy
Phys.} {\bf 0006} (2000) 021, {\tt hep-th/0005040}; R.~Gopakumar, J.~Maldacena,
S.~Minwalla and A.~Strominger, {\sl J. High Energy Phys.} {\bf 0006} (2000)
036, {\tt hep-th/0005048}.

\bibitem{Baker} G.~Baker, {\sl Phys. Rev.} {\bf 109} (1958) 2198.

\bibitem{MMVZ} V.I.~Man'ko, G.~Marmo, P.~Vitale and F.~Zaccaria, {\sl Int. J.
Mod.\ Phys.\ {\bf A9}} (1994) 5541, {\tt hep-th/9310053}.

\bibitem{Rajeev} S.G.~Rajeev, {\sl Phys.\ Rev.\ }{\bf D42} (1990) 2779; {\bf
D44} (1991) 1836.

\bibitem{CorSch} L.~Cornalba and R.~Schiappa, {\tt hep-th/0101219}.

\bibitem{NCsolitons} R.~Gopakumar, S.~Minwalla and A.~Strominger, {\sl J. High
Energy Phys.} {\bf 0005} (2000) 020, {\tt hep-th/0003160}; K.~Dasgupta,
S.~Mukhi and G.~Rajesh, {\sl J. High Energy Phys.} {\bf 0006} (2000) 022, {\tt
hep-th/0005006}; J.A.~Harvey, P.~Kraus, F.~Larsen and E.J.~Martinec, {\sl J.
High Energy Phys.} {\bf 0007} (2000) 042, {\tt hep-th/0005031}.

\bibitem{Bak} A.P. Polychronakos, Phys. Lett. {\bf B495} (2000) 407, 
{\tt hep-th/0007043}; D.~Bak, {\sl Phys. Lett.} {\bf B495} (2000) 251, {\tt
hep-th/0008204}; D.~Bak, K.~Lee and J.-H.~Park, {\sl Phys. Rev.} {\bf D63}
(2001) 125010, {\tt hep-th/0011099}.

\bibitem{fuzzysolitons} I.~Bars, H.~Kajiura, Y.~Matsuo and T.~Takayanagi, {\sl
Phys. Rev.} {\bf D63} (2001) 086001, {\tt hep-th/0010101}.


\bibitem{KTheory} R.~Minasian and G.~Moore, {\sl J. High Energy Phys.} {\bf
9711} (1997) 002, {\tt hep-th/9710230}; P.~Ho\v{r}ava, {\sl Adv. Theor. Math.
Phys.} {\bf 2} (1998) 1373, {\tt hep-th/9812135}; K.~Olsen and R.J.~Szabo, {\sl
Adv. Theor. Math. Phys.} {\bf 3} (1999) 889, {\tt hep-th/9907140}.

\bibitem{WittenK} E.~Witten, {\sl J. High Energy Phys.} {\bf 9812} (1998) 019,
{\tt hep-th/9810188}.

\bibitem{WittenNC} E.~Witten, {\tt hep-th/0006071}.

\bibitem{NCtachyons} J.A.~Harvey and G.~Moore, {\tt hep-th/0009030}.

\bibitem{twistK} A.~Kapustin, {\sl Adv. Theor. Math. Phys.} {\bf 4} (2000) 127,
{\tt hep-th/9909089}; P.~Bouwknegt and V.~Mathai, {\sl J. High Energy Phys.}
{\bf 0003} (2000) 007, {\tt hep-th/0002023}; E.~Witten, {\sl Int. J. Mod.
Phys.} {\bf A16} (2001) 693, {\tt hep-th/0007175}.

\end{thebibliography}
\end{document}